\documentclass[10pt,journal,final,twocolumn]{IEEEtran}

\usepackage{graphicx}
\usepackage{subfigure}
\usepackage{epstopdf}
\usepackage{epsfig}
\usepackage[cmex10]{amsmath}
\usepackage{amssymb}
\usepackage{times}
\usepackage{ntheorem}
\usepackage{pifont}
\usepackage{stfloats}
\usepackage{bm}
\usepackage{color}
\usepackage{makecell}
\usepackage{array}

\begin{document}
%
\title{Spatially Correlated Massive MIMO Relay Systems with Low-Resolution ADCs}


\author{\IEEEauthorblockN{Peihao~Dong,~\IEEEmembership{Student Member,~IEEE},
Hua Zhang,~\IEEEmembership{Member,~IEEE}, Qihui Wu,~\IEEEmembership{Senior Member,~IEEE} \\and Geoffrey Ye Li,~\IEEEmembership{Fellow,~IEEE}}

\vspace{-0.5cm}
\thanks{
P. Dong, H. Zhang are with the National Mobile Communications Research Laboratory, Southeast University, Nanjing 210096, China (e-mail: phdong@seu.edu.cn; huazhang@seu.edu.cn).

Q. Wu is with the College of Electronic and Information Engineering, Nanjing University of Aeronautics and Astronautics, Nanjing 211106, China (e-mail: wuqihui2014@sina.com).

G. Y. Li is with the School of Electrical and Computer Engineering, Georgia Institute of Technology, Atlanta, GA 30332 USA (e-mail: liye@ece.gatech.edu).}
}


\IEEEtitleabstractindextext{%
\begin{abstract}
In this paper, we investigate the massive MIMO relay system, where the relay station (RS) forwards the signals from multiple remote users to the base station (BS). Large-scale antenna arrays in conjunction with low-resolution analog-to-digital converters (ADCs) are equipped at the RS and the BS to guarantee the high spectral efficiency with low cost. Considering the ever-present spatial correlation at both the RS and the BS, we first study the canonical channel estimation process, from which a tractable equivalent form of the channel estimate is extracted for further analysis. Under these transmission impairments along with the ADC quantization imperfection, we derive the closed-form approximation of the achievable rate. Then the impacts of power scaling, spatial correlation level, and ADC resolution bits are revealed comprehensively to guide the practical system deployment and implementation. Numerical results are presented to verify the theoretical analysis in a straightforward way.
\end{abstract}

}

\maketitle

\IEEEdisplaynontitleabstractindextext

%
\IEEEpeerreviewmaketitle

\section{Introduction}

Massive multiple-input multiple-output (MIMO) technique has shown its dramatic capacity to improve the spectral efficiency by using usually hundreds of antennas along with efficient transceiver algorithms \cite{E. Larsson}, \cite{L. Lu}. In the era of internet of things \cite{Q. Wu_a}, \cite{GuiWCM2020}, the massive MIMO architecture is endowed new vitalities by orchestrating it with millimeter wave communications, relaying transmission, and machine learning algorithms, and thus is anticipated to bring revolutionary performance improvement \cite{A. L. Swindlehurst}--\cite{GuiTVT2018b}. Specifically, massive MIMO relaying transmission is a promising technique to enlarge the coverage and inherit the merits of massive MIMO architectures provided that both the relay station (RS) and the base station (BS) are equipped with large-scale antenna arrays. To mitigate the hardware cost and power consumption caused by massive antennas, low-resolution analog-to-digital converters (ADCs) driven massive MIMO has been extensively studied as one of the mainstream architectures for the upcoming fifth generation (5G) and future mobile networks \cite{J. Zhang_a}, \cite{S. Gao}.

\subsection{Prior Related Work}

The nonlinear signal distortion caused by low-resolution ADCs hinders the insightful performance analysis. By resorting to the additive quantization noise model (AQNM), the nonlinear ADC quantization is approximated as a linear operation with a fairly good accuracy, which enables tractable performance analysis and further yields comprehensive insights for system design \cite{J. Singh}, \cite{O. Orhan}. Based on this rule, the achievable rate expression has been derived in closed-form for multi-user (MU) massive MIMO systems with low-resolution ADCs in \cite{L. Fan}. In \cite{P. Dong_a}, the low-resolution ADC relaying has proven to be an efficient way to improve the communication quality between the remote users and the BS significantly with the low deployment cost. In \cite{C. Kong_a}, the achievable rate has been analyzed for the multi-pair massive MIMO full-duplex relay system with low-resolution ADCs at both the RS and destinations. The impacts of variable-bit ADCs and digital-to-analog converters have been studied for the multi-pair massive MIMO relay system in \cite{Y. Xiong}. In \cite{P. Dong_c}, the impact of low-resolution ADCs for MU massive MIMO systems in spatially correlated channels has been shown less severe than that in independent and identically distributed (i.i.d.) channels.

For massive MIMO relay systems, spatial correlation must be considered in performance analysis to make the extracted observations and optimization design convincing in the real scenarios. For the full-duplex two-way massive MIMO relaying system, deterministic equivalent of the achievable rate has been derived in \cite{J. Feng}, followed by the comprehensive analysis focusing on the impacts of antenna correlation at the RS and users. In \cite{D. N. Amudala}, multiple source and destination users have been further considered and the lower bound of spectral efficiency has been derived, based on which the optimal power control is proposed. The hybrid transceiver architecture has been integrated into the multi-pair massive MIMO relay system and the analog processing matrices have been optimized based on the derived spectral efficiency expressions in \cite{M. Fozooni}.

\subsection{Contribution}

From the perspective of practical implementation, it is necessary to consider both the transmission impairments, i.e., spatial correlation and imperfect channel state information (CSI), and the ADC quantization error when analyzing the massive MIMO relay systems. However, such comprehensive research considering these practical factors is still missing in the literature. In this paper, we fill this gap to provide insightful guidance for system deployment and design. For the considered system, the RS equipped with massive antennas enables the transmission from the remote users to the BS equipped with massive antennas. Dedicated low-resolution ADC is used for each antenna at the RS and the BS to reduce the deployment and operating costs. The main contribution of this paper can be summarized as follows.
\begin{itemize}[\IEEEsetlabelwidth{Z}]
\item[1)] To acquire the CSI for signal detection, we first study the canonical linear minimum mean-squared error (LMMSE) channel estimation for the two-hop channels in presence of spatial correlation and ADC quantization error. Then a tractable equivalent form of the channel estimate is figured out to pave the following achievable rate analysis.

\item[2)] Under the transmission impairments and ADC quantization imperfection, a tight closed-form approximation of the achievable sum rate with maximal ratio combining (MRC) detection is derived.\footnote{For massive MIMO systems, MRC is a computationally efficient detection technique in two-fold aspects: 1) It avoids the intractable high-dimensional matrix inversion by simply multiplying the received signals with conjugate channel coefficients; 2) It can be implemented in the decentralized manner at each antenna unit, which reduces the overhead for fronthaul data transmission significantly and further accelerates the computation. Therefore, MRC can maintain the low cost and power consumption and thus is widely studied in massive MIMO relay systems \cite{P. Dong_a}, \cite{D. N. Amudala}, \cite{C. Kong_b}.} The challenge mainly originates from the cascaded two-hop channels with double-side spatial correlation and the effect of imperfect ADC quantization, in which case the widely used derivation methods for the performance analysis of massive MIMO systems cannot be applied directly. We address this problem by proposing an important preliminary lemma, which greatly facilitates the derivation and can be also applied to other similar problems.
    %

\item[3)] Based on the derived approximated expression of the sum rate, the general power scaling law is extracted, which demonstrates that the spatial correlation does not impact the asymptotic behavior of the achievable rate with power scaling. Further analysis reveals that the impacts of spatial correlation and ADC resolution on the performance are dependent on the relationship between the numbers of antennas at the RS and the BS. These insights provide straightforward guidance for the practical system deployment.
\end{itemize}

\subsection{Difference from Related Work}

Compared to the related works, e.g., \cite{P. Dong_a}--\cite{C. Kong_c}, the cascaded two-hop channels with double-side spatial correlation, imperfect CSI, and ADC quantization error considered in this paper pose an unprecedented challenge in the achievable rate analysis of massive MIMO relay systems and require proper innovations in the analytical approach. To address this challenge, the novelty of this paper can be summarized as the following two aspects:
\begin{itemize}[\IEEEsetlabelwidth{Z}]
\item[1)] Different from most prior works, the results of LMMSE channel estimation cannot be directly applied to the subsequent achievable rate analysis and are equivalently transformed to the tractable forms. Although \cite{M. Fozooni} handles the problem in the similar way, it only considers the single-side spatial correlation and the ADC quantization error is not incorporated.

\item[2)] With the double-side spatial correlation, channels become correlated instead of mutually orthogonal and the analytical approaches for i.i.d. channels and single-side correlated channels malfunction. The imperfect CSI and ADC quantization error further complicate the achievable rate analysis. To tackle this problem, we propose a novel and general lemma to reveal the statistical property of the double-side correlated channels. With this lemma and the equivalent forms of the channel estimates, the analytical approach in our previous work \cite{P. Dong_a} can be applied to yield a closed-form approximation of the achievable rate, based on which we know how to deploy the RS and the BS practically.
\end{itemize}

The rest of the paper is organized as follows. Section II describes the considered massive MIMO relay system including channel model and signal transmission process. Section III studies LMMSE channel estimation for the considered system, based on which Section IV derives the closed-form approximated expression of the achievable sum rate with MRC detection. The impacts of power scaling, spatial correlation, and ADC resolution are then extracted. Numerical results are provided in Section V to validate the theoretical analysis and finally Section VI concludes this paper.

\emph{Notations}: In this paper, we use upper and lower case boldface letters to denote matrices and vectors, respectively. $(\cdot)^T$, $(\cdot)^*$, $(\cdot)^H$, $(\cdot)^{-1}$, $\text{tr}(\cdot)$ and $\mathbb{E}\{\cdot\}$ represent the transpose, conjugate, conjugate transpose, inverse, trace, and expectation, respectively. $\|\cdot\|$ denotes the Euclidean norm of a vector. $\|\cdot\|_2$ and $\|\cdot\|_F$ denote the spectral norm and Frobenius norm of a matrix, respectively. $\otimes$ denotes the Kronecker product. $\mathcal{CN}(\mu,\sigma^2)$ represents a circularly symmetric complex Gaussian distribution with mean $\mu$ and variance $\sigma^2$. $\mathbf{I}_N$ denotes an $N\times N$ identify matrix. $f(x)=\mathcal{O}(g(x))$ means that $0<\lim\limits_{x \to \infty}\frac{f(x)}{g(x)}<\infty$.

\section{System Model}

After briefly introducing channel model, we discuss signal transmission process in this section.

\subsection{Channel Model}

\begin{figure}[!t]
\centering
\includegraphics[trim=0 0 0 0, width=2.5in]{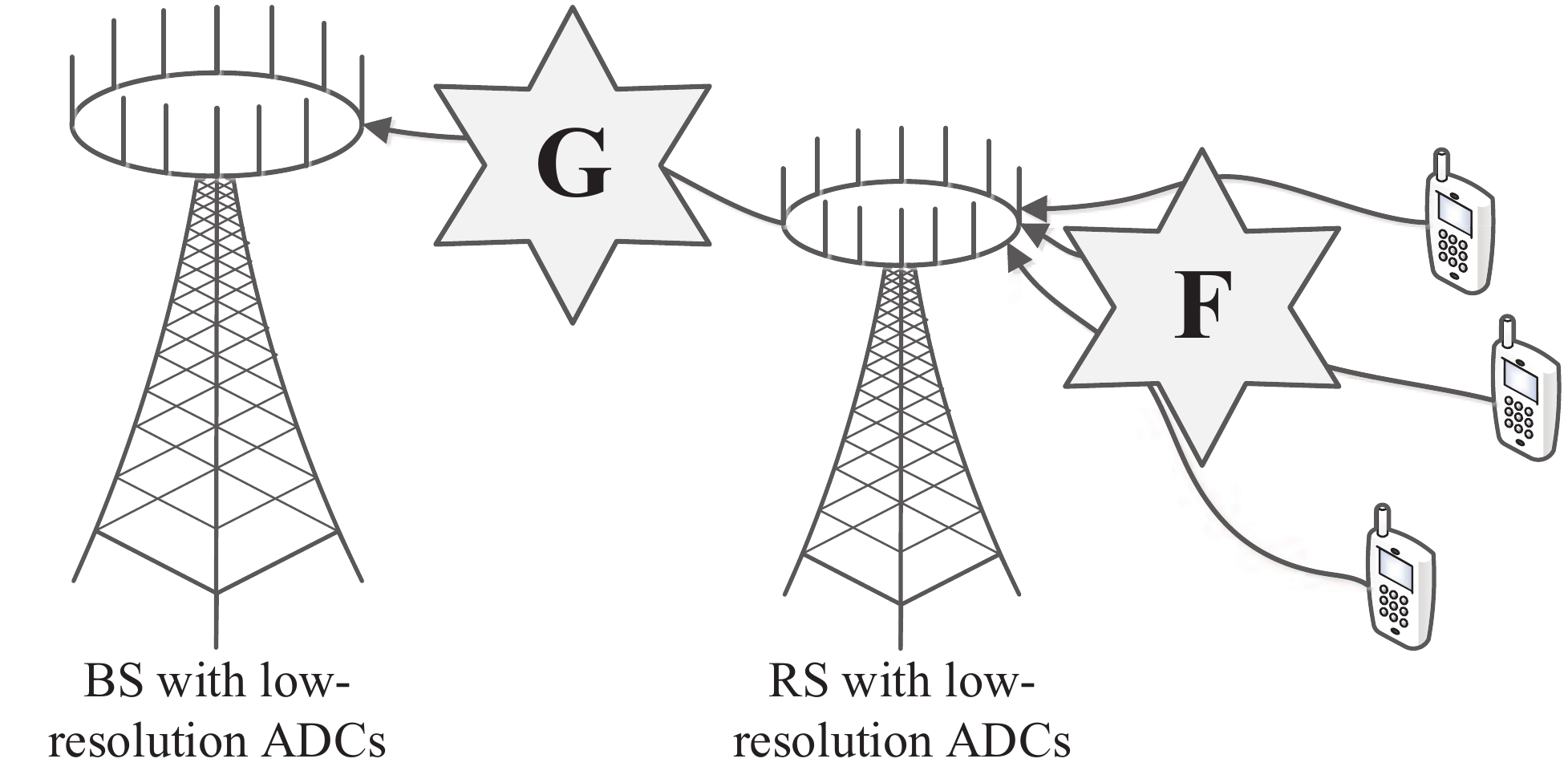}
\caption{System model of a massive MIMO relay system with low-resolution ADCs.}\label{system_model}
\end{figure}

Fig.~\ref{system_model} shows a MU massive MIMO relay uplink, where $K$ single-antenna users communicate with a BS with $M$ antennas via an amplify-and-forward (AF) half-duplex RS with $N$ antennas.\footnote{According to \cite{X. Meng}, the time synchronization among all users is realized with the aid of synchronization signals broadcasted by the RS. The carrier frequency offsets can be well estimated and compensated by the frequency synchronization algorithm \cite{W. Zhang}.} Both the BS and the RS are equipped with large numbers of antennas ($M,N\gg 1$) and low-resolution ADC for each antenna to guarantee simultaneous services for an arbitrary number of remote users as well as to reduce the transmit power of each user in the low cost. Assume that there is no direct links from the users to the BS due to heavy shadowing and path loss. Spatial correlation is considered at both the RS and the BS and is modeled by the Kronecker model \cite{D.-S. Shiu}. Specifically, $\mathbf{F}\in\mathbb{C}^{N\times K}$ denotes the channel matrix from all $K$ users to the RS. It can be decomposed into  $\mathbf{F}=\mathbf{T}_{\textrm{R}_{\textrm{r}}}^{\frac{1}{2}}\mathbf{H}_{\textrm{F}}\mathbf{D}_{\textrm{F}}^{\frac{1}{2}}$, where $\mathbf{T}_{\textrm{R}_{\textrm{r}}}\in\mathbb{C}^{N\times N}$ denotes the receive correlation matrix at the RS, $\mathbf{H}_{\textrm{F}}\in\mathbb{C}^{N\times K}$, with i.i.d. $\mathcal{CN}(0,1)$ elements, represents the small-scale fading, and the diagonal matrix $\mathbf{D}_{\textrm{F}}\in\mathbb{C}^{K\times K}$ accounts for the large-scale fading with the $k$th diagonal element denoted by $\beta_k$. For the second hop, $K$ RS antennas are selected to forward the amplified signal to the BS \cite{P. Dong_a} and the channel matrix $\mathbf{G}\in\mathbb{C}^{M\times K}$ can be written as $\mathbf{G}=\sqrt{\eta}\mathbf{T}_{\textrm{B}_{\textrm{r}}}^{\frac{1}{2}}\mathbf{H}_{\textrm{G}}\mathbf{T}_{\textrm{R}_{\textrm{t}}}^{\frac{1}{2}}$, where $\mathbf{T}_{\textrm{B}_{\textrm{r}}}\in\mathbb{C}^{M\times M}$ and $\mathbf{T}_{\textrm{R}_{\textrm{t}}}\in\mathbb{C}^{K\times K}$ denote the receive correlation matrix at the BS and the transmit correlation matrix at the RS, $\eta$ and $\mathbf{H}_{\textrm{G}}\in\mathbb{C}^{M\times K}$ represent the corresponding large-scale fading coefficient and small-scale fading matrix with i.i.d. $\mathcal{CN}(0,1)$ elements, respectively. The spatial correlation matrices $\mathbf{T}_{\textrm{R}_{\textrm{r}}}$, $\mathbf{T}_{\textrm{R}_{\textrm{t}}}$, and $\mathbf{T}_{\textrm{B}_{\textrm{r}}}$ are positive semi-definite Hermitian with uniformly bounded spectrum norms and unit diagonal elements \cite{J. Feng}, \cite{C. Martin}.


\subsection{Signal Transmission}

\begin{table}[!t]
\renewcommand{\arraystretch}{1.3}
\centering
\caption{Distortion Factor $\rho_1$ vs. ADC Resolution Bit $q_1$}
\label{table_1}
\begin{tabular}{c|c|c|c|c|c}
\hline
$q_1$ & 1 & 2 & 3 & 4 & 5\\
\hline
$\rho_1$ & 0.3634 & 0.1175 & 0.03454 & 0.009497 & 0.002499\\
\hline
\end{tabular}
\end{table}

Each coherence interval is halved into two time slots for the original signal transmission from $K$ users to the RS and the retransmission from the RS to the BS, respectively. In the first time slot, the received signal at the RS is given by
\begin{eqnarray}
\label{eqn_yR}
\mathbf{y}_{\textrm{R}}=\sqrt{P_{\textrm{U}}}\mathbf{F} \mathbf{x}_{\textrm{U}}+\mathbf{n}_{\textrm{R}},
\end{eqnarray}
where $P_{\textrm{U}}$ denotes the transmit power for each user, $\mathbf{x}_{\textrm{U}}=\left[x_{\textrm{U},1}, \dots, x_{\textrm{U},K}\right]^T$ includes the transmit signals of all $K$ users with $\mathbb{E}\{\mathbf{x}_{\textrm{U}}\mathbf{x}^H_{\textrm{U}}\}=\mathbf{I}_K$, and $\mathbf{n}_{\textrm{R}}$ denotes the additive white Gaussian noise (AWGN) at the RS with i.i.d. $\mathcal{CN}(0, \sigma_{\textrm{R}}^2)$ elements. Before the digital processing, $\mathbf{y}_{\textrm{R}}$ is first quantized by the low-resolution ADCs at the RS as ${\mathbf{y}}_{\text{R,q}}=\mathbb{Q}\left(\mathbf{y}_{\textrm{R}}\right)$, where $\mathbb{Q}\left(\cdot\right)$ denotes the nonlinear quantization operation on the real and imaginary parts in the element-wise manner. To facilitate further analysis, we resort to the widely used linear AQNM to expand ${\mathbf{y}}_{\text{R,q}}$ as \cite{O. Orhan}, \cite{A. K. Fletcher}
\begin{eqnarray}
\label{eqn_yRq}
\mathbf{y}_{\text{R,q}}=\alpha_1\mathbf{y}_{\textrm{R}}+\mathbf{n}_{\text{q}1}=\alpha_1\sqrt{P_{\textrm{U}}}\mathbf{F} \mathbf{x}_{\textrm{U}}+\alpha_1\mathbf{n}_{\textrm{R}}+\mathbf{n}_{\text{q}1},
\end{eqnarray}
where $\alpha_1=1-\rho_1$ is the linear quantization gain with $\rho_1=\frac{\mathbb{E}\{\|\mathbf{y}_{\textrm{R}}-\mathbf{y}_{\text{R,q}}\|^2\}}{\mathbb{E}\{\|\mathbf{y}_{\textrm{R}}\|^2\}}$ denoting the distortion factor caused by low-resolution ADCs at the RS. According to \cite{Q. Bai}, the exact values of $\rho_1$, dependent on the ADC resolution, are listed in Table~\ref{table_1} for the resolution bit $q_1\leq5$ and can be approximated by $\rho_1=\frac{\sqrt{3}\pi}{2}\cdot 2^{-2q_1}$ for $q_1\geq6$. $\mathbf{n}_{\text{q}1}$ is the quantization noise uncorrelated with $\mathbf{y}_{\textrm{R}}$. We assume the worst case that the quantization noise $\mathbf{n}_{\text{q}1}$ is Gaussian distributed, which provides a tractable lower bound on the achievable rate. Specifically, the conditional covariance matrix of $\mathbf{n}_{\text{q}1}$ under a fixed channel realization $\mathbf{F}$ is given by
\begin{eqnarray}
\label{eqn_Rnq}
\mathbf{R}_{\mathbf{n}_{\text{q}1}}=\alpha_1\rho_1\text{diag}(P_{\textrm{U}} \mathbf{F} \mathbf{F}^H+\sigma_{\textrm{R}}^2\mathbf{I}_{N}).
\end{eqnarray}

After ADC quantization, MRC is applied by multiplying ${\mathbf{y}}_{\text{R,q}}$ with the detection matrix, $\mathbf{W}^H$, that depends on $\mathbf{F}$ as
\setlength{\arraycolsep}{0.0em}
\begin{eqnarray}
\label{eqn_rR}
\mathbf{r}_{\textrm{R}}&&=\mathbf{W}^H{\mathbf{y}}_{\text{R,q}}\nonumber\\
&&=\alpha_1\sqrt{P_{\textrm{U}}}\mathbf{W}^H\mathbf{F} \mathbf{x}_{\textrm{U}}+\alpha_1\mathbf{W}^H\mathbf{n}_{\textrm{R}}+\mathbf{W}^H\mathbf{n}_{\text{q}1}.
\end{eqnarray}
Then the signal retransmitted by the RS is written as
\begin{eqnarray}
\label{eqn_xR}
\mathbf{x}_{\textrm{R}}=\kappa\mathbf{r}_{\textrm{R}},
\end{eqnarray}
where $\kappa$ is the amplification factor imposed to satisfy a long-term total transmit power constraint at the RS, i.e. $\mathbb{E}\{\mathbf{x}^H_{\textrm{R}}\mathbf{x}_{\textrm{R}}\}=P_{\textrm{R}}$, and is given by (\ref{eqn_kappa}), shown at the top of the next page.

\newcounter{mytempeqncnt}
\begin{figure*}[!t]
\normalsize
\setcounter{mytempeqncnt}{\value{equation}}
\setcounter{equation}{5}
\begin{equation}
\label{eqn_kappa}
\kappa=\sqrt{\frac{P_{\textrm{R}}}{\alpha_1^2P_{\textrm{U}} \text{tr}\left(\mathbb{E}\left\{\mathbf{W}^H\mathbf{F}\mathbf{F}^H\mathbf{W}\right\}\right)+\alpha_1(1-\alpha_1)P_{\textrm{U}} \text{tr}\left(\mathbb{E}\left\{\mathbf{W}^H\text{diag}(\mathbf{F} \mathbf{F}^H)\mathbf{W}\right\}\right)+\alpha_1\sigma_{\textrm{R}}^2 \text{tr}\left(\mathbb{E}\left\{\mathbf{W}^H\mathbf{W}\right\}\right)}}.
\end{equation}
\setcounter{equation}{\value{mytempeqncnt}}
\hrulefill
\vspace*{-6pt}
\end{figure*}

In the second time slot, the RS selects $K$ antennas to forward $\mathbf{x}_{\textrm{R}}$. The received signal at the BS is expressed as
\setcounter{equation}{6}
\begin{equation}
\label{eqn_yB}
\mathbf{y}_{\textrm{B}}=\mathbf{G} \mathbf{x}_{\textrm{R}}+\mathbf{n}_{\textrm{B}},
\end{equation}
where $\mathbf{n}_B\sim\mathcal{CN}(0, \sigma_{\textrm{B}}^2 \mathbf{I}_{M})$ is the AWGN at the BS. Similar to the quantization of $\mathbf{y}_{\textrm{R}}$ at the RS, $\mathbf{y}_{\textrm{B}}$ is quantized by the low-resolution ADCs at the BS before further processing. Then the approximated quantization output by AQNM is given by
\begin{eqnarray}
\label{eqn_yBq}
\mathbf{y}_{\text{B,q}}=\alpha_2\mathbf{y}_{\textrm{B}}+\mathbf{n}_{\text{q}2}=\alpha_2\mathbf{G} \mathbf{x}_{\textrm{R}}+\alpha_2\mathbf{n}_{\textrm{B}}+\mathbf{n}_{\text{q}2},
\end{eqnarray}
where $\alpha_2=1-\rho_2$ is the linear quantization gain dependent on the ADC resolution bit $q_2$ at the BS and the corresponding value of $\rho_2$ is taken similarly to $\rho_1$. The quantization noise $\mathbf{n}_{\text{q}2}$ is uncorrelated with $\mathbf{y}_{\textrm{B}}$ and is also assumed Gaussian distributed with its conditional covariance matrix under a fixed channel realization $\mathbf{G}$ approximated by
\begin{eqnarray}
\label{eqn_Rnq}
\mathbf{R}_{\mathbf{n}_{\text{q}2}}\approx\alpha_2\rho_2\text{diag}\left(\frac{P_{\textrm{R}}}{K} \mathbf{G} \mathbf{G}^H+\sigma_{\textrm{B}}^2\mathbf{I}_{M}\right).
\end{eqnarray}

Finally, the MRC detection matrix, $\mathbf{A}^H$, that depends on $\mathbf{G}$ is used to multiply ${\mathbf{y}}_{\text{B,q}}$ as
\begin{eqnarray}
\label{eqn_rB}
\mathbf{u}_{\textrm{B}}&&=\mathbf{A}^H\mathbf{y}_{\textrm{B,q}}\nonumber\\
&&=\alpha_{1}\alpha_{2}\kappa\sqrt{P_{\textrm{U}}}\mathbf{A}^H\mathbf{G}\mathbf{W}^H\mathbf{F}\mathbf{x}_{\textrm{U}} +\alpha_{1}\alpha_{2}\kappa\mathbf{A}^H\mathbf{G}\mathbf{W}^H\mathbf{n}_{\textrm{R}}\nonumber\\
&&+\alpha_{2}\kappa\mathbf{A}^H\mathbf{G}\mathbf{W}^H\mathbf{n}_{\textrm{q}1} +\alpha_{2}\mathbf{A}^H\mathbf{n}_{\textrm{B}}+\mathbf{A}^H\mathbf{n}_{\textrm{q}2},
\end{eqnarray}
where the $k$th element of $\mathbf{u}_{\textrm{B}}$ is used to detect the signal originating from the $k$th user.

\section{Channel Estimation}

In this section, the canonical LMMSE channel estimation is first studied, based on which the tractable estimate forms of the channel matrices, $\mathbf{F}$ and $\mathbf{G}$, are derived, respectively, by considering both the spatial correlation and ADC quantization at the RS and the BS.

\subsection{Estimation of $\mathbf{F}$}

To estimate $\mathbf{F}$ at the RS, all users simultaneously transmit mutually orthogonal pilot sequences of length $\tau_1$ ($\geq K$) symbols, which is denoted by a $\tau_1\times K$ matrix $\sqrt{\tau_1}\boldsymbol{\Phi}$ with $\boldsymbol{\Phi}^H \boldsymbol{\Phi}=\mathbf{I}_K$. The received pilots at the RS during $\tau_1$ time instants are given by
\begin{eqnarray}
\label{eqn_SR}
\mathbf{S}_{\text{R}}=\sqrt{\tau_1 P_1}\mathbf{F}\boldsymbol{\Phi}^T+\mathbf{N}_{\text{R}_{\textrm{p}}},
\end{eqnarray}
where $P_1$ denotes the transmit power of each user during a pilot symbol period and $\mathbf{N}_{\text{R}_{\textrm{p}}}\in \mathbb{C}^{N\times\tau_1}$ denotes AWGN during the pilot transmission phase. $\mathbf{S}_{\text{R}}$ will be quantized
by the low-resolution ADCs at the RS, yielding
\setlength{\arraycolsep}{0.05em}
\begin{eqnarray}
\label{eqn_Sq}
\mathbf{S}_{\text{R,q}}&&=\alpha_{1}\mathbf{S}_{\text{R}}+\mathbf{N}_{\text{p,q}1}\nonumber\\
&&=\alpha_{1}\sqrt{\tau_1 P_1}\mathbf{F}\boldsymbol{\Phi}^T+\alpha_{1}\mathbf{N}_{\text{R}_{\textrm{p}}}+\mathbf{N}_{\text{p,q}1},
\end{eqnarray}
where $\mathbf{N}_{\text{p,q}1}$ is the quantization noise uncorrelated with $\mathbf{S}_{\text{R}}$. By postmultiplying $\mathbf{S}_{\text{R,q}}$ with $\boldsymbol{\Phi}^{*}$, we have
\setlength{\arraycolsep}{0.2em}
\begin{eqnarray}
\label{eqn_ZR}
\mathbf{Z}_{\text{R}}=\mathbf{S}_{\text{R,q}}\boldsymbol{\Phi}^{*}=\alpha_{1}\sqrt{\tau_1 P_1}\mathbf{F}+\alpha_{1}\mathbf{N}_{\text{R}_{\textrm{p}}}\boldsymbol{\Phi}^{*}+\mathbf{N}_{\text{p,q}1}\boldsymbol{\Phi}^{*},
\end{eqnarray}
based on which the LMMSE estimate of $\mathbf{F}$ is given by
\begin{eqnarray}
\label{eqn_hatF}
\hat{\mathbf{F}}=\mathbf{Q}_{\textrm{F}}\mathbf{Z}_{\text{R}}.
\end{eqnarray}
Then the MSE between $\hat{\mathbf{F}}$ and $\mathbf{F}$ is given by
\setlength{\arraycolsep}{0.0em}
\begin{eqnarray}
\label{eqn_MSE}
\mathbb{E}\{\|\hat{\mathbf{F}}-\mathbf{F}\|^2_F\} &&=\textrm{tr}\left(\mathbb{E}\{(\hat{\mathbf{F}}-\mathbf{F})(\hat{\mathbf{F}}-\mathbf{F})^H\}\right)\nonumber\\
&&=\textrm{tr}(\mathbf{Q}_{\textrm{F}}\mathbf{R}_{\textrm{Z}_{\textrm{R}}}\mathbf{Q}_{\textrm{F}}^H) +N\sum_{i=1}^K\beta_i\nonumber\\
&&-\alpha_{1}\sqrt{\tau_1 P_1}\sum_{i=1}^K\beta_i\textrm{tr}(\mathbf{Q}_{\textrm{F}}\mathbf{T}_{\textrm{R}_{\textrm{r}}} +\mathbf{T}_{\textrm{R}_{\textrm{r}}}\mathbf{Q}_{\textrm{F}}^H),\quad
\end{eqnarray}
where
\setlength{\arraycolsep}{0.0em}
\begin{eqnarray}
\label{eqn_RzR}
\mathbf{R}_{\textrm{Z}_{\textrm{R}}}&&=\mathbb{E}\{\mathbf{Z}_{\text{R}}\mathbf{Z}_{\text{R}}^H\} =\alpha_1^2\tau_1P_1\sum_{i=1}^K\beta_i\mathbf{T}_{\textrm{R}_{\textrm{r}}}\nonumber\\
&&\quad+K\alpha_1\left((1-\alpha_1)P_1\sum_{i=1}^K\beta_i+\sigma_R^2\right)\mathbf{I}_N.
\end{eqnarray}
We next compute the derivative of $\mathbb{E}\{\|\hat{\mathbf{F}}-\mathbf{F}\|^2_F\}$ with respect to $\mathbf{Q}_{\textrm{F}}^*$ and force it to be $\mathbf{0}$, that is
\begin{eqnarray}
\label{eqn_Q}
\frac{\partial \mathbb{E}\{\|\hat{\mathbf{F}}-\mathbf{F}\|^2_F\}}{\partial \mathbf{Q}_{\textrm{F}}^*} =\mathbf{Q}_{\textrm{F}}\mathbf{R}_{\textrm{Z}_{\textrm{R}}}-\alpha_{1}\sqrt{\tau_1 P_1}\sum_{i=1}^K\beta_i\mathbf{T}_{\textrm{R}_{\textrm{r}}}=\mathbf{0},\quad
\end{eqnarray}
which leads to
\begin{eqnarray}
\label{eqn_Q}
\mathbf{Q}_{\textrm{F}}=\alpha_1\sqrt{\tau_1P_1}\sum_{i=1}^K\beta_i\mathbf{T}_{\textrm{R}_{\textrm{r}}}\mathbf{R}_{\textrm{Z}_{\textrm{R}}}^{-1}.
\end{eqnarray}
Then the the LMMSE estimate of $\mathbf{F}$ can be obtained by substituting (\ref{eqn_Q}) into (\ref{eqn_hatF}). The MSE in (\ref{eqn_MSE}) can be also compacted as
\begin{eqnarray}
\label{eqn_MSE2}
\mathbb{E}\{\|\hat{\mathbf{F}}\!-\!\mathbf{F}\|^2_F\}&&=\!\sum_{i=1}^K\beta_i\!\left(\!N\!-\!\alpha_{1}^2\tau_1 P_1\sum_{i=1}^K\beta_i\textrm{tr}(\mathbf{T}_{\textrm{R}_{\textrm{r}}}\mathbf{R}_{\textrm{Z}_{\textrm{R}}}^{-1}\mathbf{T}_{\textrm{R}_{\textrm{r}}})\!\! \right)\nonumber\\
&&=\sum_{i=1}^K\beta_i\!\left(\!N\!-\!\alpha_{1}^2\tau_1 P_1\sum_{i=1}^K\beta_i\sum_{n=1}^N\frac{\lambda_{\textrm{R}_{\textrm{r}},n}^2}{\Lambda_1}\!\! \right),
\end{eqnarray}
where $\Lambda_1\!=\!\alpha_1^2\tau_1P_1\sum_{i=1}^K\beta_i\lambda_{\textrm{R}_{\textrm{r}},n} \!+\!K\alpha_1((1\!-\!\alpha_1)P_1\sum_{i=1}^K\beta_i\!+\!\sigma_R^2)$ with $\lambda_{\textrm{R}_{\textrm{r}},n}$ denoting the $n$th largest eigenvalue of $\mathbf{T}_{\textrm{R}_{\textrm{r}}}$. (\ref{eqn_MSE2}) reveals that the estimation error of $\mathbf{F}$ decreases with $\alpha_1$, i.e., the ADC resolution at the RS. In addition, the imperfect ADC quantization at the RS leads to a non-vanishing estimation error floor for $\mathbf{F}$ when $P_1\rightarrow\infty$.

The expression of $\hat{\mathbf{F}}$ in (\ref{eqn_hatF}) is intractable for further analysis. To address this problem, we resort to \cite[Eq. (12)]{M. Fozooni} and rewrite $\hat{\mathbf{F}}$ as
\begin{eqnarray}
\label{eqn_hatF_trac_form}
\hat{\mathbf{F}}=\left(\frac{\mathbb{E}\{\hat{\mathbf{F}}\hat{\mathbf{F}}^H\}} {\textrm{tr}\left(\mathbb{E}\{\hat{\mathbf{F}}^H\hat{\mathbf{F}}\}\right)}\right)^{\frac{1}{2}}\hat{\mathbf{H}}_{\textrm{F}} \left(\mathbb{E}\{\hat{\mathbf{F}}^H\hat{\mathbf{F}}\}\right)^{\frac{1}{2}},
\end{eqnarray}
where $\hat{\mathbf{H}}_{\textrm{F}}\in\mathbb{C}^{N\times K}$ has i.i.d. $\mathcal{CN}(0,1)$ elements. The equality in (\ref{eqn_hatF_trac_form}) holds since the both sides have the same mean value and second-order expectation. By deriving the expectations, the equivalent form of $\hat{\mathbf{F}}$ in (\ref{eqn_hatF_trac_form}) can be further compacted as
\begin{eqnarray}
\label{eqn_hatF_trac_form2}
\hat{\mathbf{F}}=\hat{\mathbf{T}}_{\textrm{R}_{\textrm{r}}}^{\frac{1}{2}}\hat{\mathbf{H}}_{\textrm{F}} \hat{\mathbf{D}}_{\textrm{F}}^{\frac{1}{2}},
\end{eqnarray}
where
\begin{eqnarray}
\label{eqn_hat_TRr}
\hat{\mathbf{T}}_{\textrm{R}_{\textrm{r}}}=\alpha_1^2\tau_1P_1\sum_{i=1}^K\beta_i\mathbf{T}_{\textrm{R}_{\textrm{r}}} \mathbf{R}_{\textrm{Z}_{\textrm{R}}}^{-1}\mathbf{T}_{\textrm{R}_{\textrm{r}}},
\end{eqnarray}
\vspace{-0.3cm}
\begin{eqnarray}
\label{eqn_hat_DF}
\hat{\mathbf{D}}_{\textrm{F}} =\frac{\textrm{tr}\left(\mathbf{D}_{\textrm{F}}\right)}{\textrm{tr}\left(\bar{\mathbf{D}}_{\textrm{F}}\right)}\bar{\mathbf{D}}_{\textrm{F}},
\end{eqnarray}
with
\begin{eqnarray}
\label{eqn_bar_DF}
\bar{\mathbf{D}}_{\textrm{F}}&&=\alpha_1^2\tau_1P_1\biggl(\sum_{i=1}^K\beta_i\biggr)^2 \biggl[\alpha_1^2\tau_1P_1 \textrm{tr}(\mathbf{T}_{\textrm{R}_{\textrm{r}}}\mathbf{R}_{\textrm{Z}_{\textrm{R}}}^{-1}\mathbf{T}_{\textrm{R}_{\textrm{r}}}^2\mathbf{R}_{\textrm{Z}_{\textrm{R}}}^{-1})\mathbf{D}_{\textrm{F}}\nonumber\\ &&\quad+\alpha_1\biggl((1-\alpha_1)P_1\sum_{i=1}^K\beta_i+\sigma_R^2\biggr) \textrm{tr}(\mathbf{R}_{\textrm{Z}_{\textrm{R}}}^{-1}\mathbf{T}_{\textrm{R}_{\textrm{r}}}^2\mathbf{R}_{\textrm{Z}_{\textrm{R}}}^{-1})\mathbf{I}_K\biggr].\nonumber\\
&&
\end{eqnarray}

Similarly, we can also obtain the channel estimation error matrix as
\begin{eqnarray}
\label{eqn_tildeF_trac_form2}
\tilde{\mathbf{F}}=\mathbf{F}-\hat{\mathbf{F}}\triangleq\tilde{\mathbf{T}}_{\textrm{R}_{\textrm{r}}}^{\frac{1}{2}}\tilde{\mathbf{H}}_{\textrm{F}} \tilde{\mathbf{D}}_{\textrm{F}}^{\frac{1}{2}},
\end{eqnarray}
where $\tilde{\mathbf{H}}_{\textrm{F}}\in\mathbb{C}^{N\times K}$ has i.i.d. $\mathcal{CN}(0,1)$ elements, $\tilde{\mathbf{T}}_{\textrm{R}_{\textrm{r}}}=\mathbf{T}_{\textrm{R}_{\textrm{r}}}-\hat{\mathbf{T}}_{\textrm{R}_{\textrm{r}}}$ and $\tilde{\mathbf{D}}_{\textrm{F}} =\frac{\textrm{tr}\left(\mathbf{D}_{\textrm{F}}\right)}{\textrm{tr}\left(\check{\mathbf{D}}_{\textrm{F}}\right)}\check{\mathbf{D}}_{\textrm{F}}$ with $\check{\mathbf{D}}_{\textrm{F}}=N\mathbf{D}_{\textrm{F}}-\bar{\mathbf{D}}_{\textrm{F}}$.

\subsection{Estimation of $\mathbf{G}$}

To estimate $\mathbf{G}$, the selected $K$ RS antennas simultaneously transmit mutually orthogonal pilot sequences of length $\tau_2$ ($\geq K$) symbols, which is denoted by a $\tau_2\times K$ matrix $\sqrt{\tau_2}\boldsymbol{\Theta}$ with $\boldsymbol{\Theta}^H \boldsymbol{\Theta}=\mathbf{I}_K$ and $P_2$ being the total transmit power of the RS during a pilot symbol period.\footnote{$\mathbf{G}$ is estimated at the BS due to two-fold reasons: 1) The BS usually has the better computation capacity than the RS. It is natural to execute the LMMSE estimation of $\mathbf{G}$ involving high-dimensional matrix inversion at the BS to relieve the computation load at the RS. This can reduce the hardware cost and power consumption for the RS and thus conforms to the rule for practical implementation. 2) It can save the pilot overhead significantly since using only $K$ time instants for pilot transmission is sufficient to obtain the reliable estimate of $\mathbf{G}$.} The estimation process is similar to that of $\mathbf{F}$ and thus the canonical LMMSE estimate of $\mathbf{G}$ is given by
\setlength{\arraycolsep}{0.2em}
\begin{eqnarray}
\hat{\mathbf{G}}=\mathbf{Q}_{\textrm{G}}\mathbf{Z}_{\text{B}}=\alpha_2\sqrt{\tau_2P_2 K} \eta \mathbf{T}_{\textrm{B}_{\textrm{r}}}\mathbf{R}_{\textrm{Z}_{\textrm{B}}}^{-1}\mathbf{Z}_{\text{B}},
\end{eqnarray}
where
\begin{eqnarray}
\mathbf{R}_{\textrm{Z}_{\textrm{B}}}\!=\!\alpha_2^2\tau_2P_2\eta\mathbf{T}_{\textrm{B}_{\textrm{r}}}\! +\!K\alpha_2\left((1\!-\!\alpha_2)P_2 \eta+\sigma_B^2\right)\mathbf{I}_M,
\end{eqnarray}
\setlength{\arraycolsep}{0.25em}
\begin{eqnarray}
\mathbf{Z}_{\textrm{B}}=\alpha_2\sqrt{\tau_2\frac{P_2}{K}}\mathbf{G} +\alpha_{2}\mathbf{N}_{\text{B}_{\textrm{p}}}\boldsymbol{\Theta}^{*}+\mathbf{N}_{\text{p,q}2}\boldsymbol{\Theta}^{*}
\end{eqnarray}
with $\mathbf{N}_{\text{B}_{\textrm{p}}}$ and $\mathbf{N}_{\text{p,q}2}$ denoting AWGN and quantization noise at the BS during the pilot transmission phase. Then the MSE between $\hat{\mathbf{G}}$ and $\mathbf{G}$ can be expressed as
\setlength{\arraycolsep}{0.0em}
\begin{eqnarray}
\label{eqn_MSE_G}
\mathbb{E}\{\|\hat{\mathbf{G}}-\mathbf{G}\|^2_F\}
=K\eta\left(\!M\!-\!\alpha_{2}^2\tau_2 P_2\eta\sum_{m=1}^M\frac{\lambda_{\textrm{B}_{\textrm{r}},m}^2}{\Lambda_2}\!\! \right),
\end{eqnarray}
where $\Lambda_2\!=\!\alpha_2^2\tau_2P_2\eta\lambda_{\textrm{B}_{\textrm{r}},m} \!+\!K\alpha_2((1-\alpha_2)P_2 \eta+\sigma_B^2)$ with $\lambda_{\textrm{B}_{\textrm{r}},m}$ denoting the $m$th largest eigenvalue of $\mathbf{T}_{\textrm{B}_{\textrm{r}}}$. From (\ref{eqn_MSE_G}), the estimation error of $\mathbf{G}$ decreases with $\alpha_2$, i.e., the ADC resolution at the BS, and is independent of the transmit correlation at the RS. The imperfect ADC quantization at the BS also causes a non-vanishing estimation error floor for $\mathbf{G}$ when $P_2\rightarrow\infty$.

Similar to the operation on $\hat{\mathbf{F}}$ in (\ref{eqn_hatF_trac_form}), the equivalent form of $\hat{\mathbf{G}}$ is expressed as
\setlength{\arraycolsep}{0.2em}
\begin{eqnarray}
\hat{\mathbf{G}}=\sqrt{\eta}\hat{\mathbf{T}}_{\textrm{B}_{\textrm{r}}}^{\frac{1}{2}}\hat{\mathbf{H}}_{\textrm{G}} \left(\frac{K}{\textrm{tr}\left(\bar{\mathbf{T}}_{\textrm{R}_{\textrm{t}}}\right)}\bar{\mathbf{T}}_{\textrm{R}_{\textrm{t}}}\right)^{\frac{1}{2}} \triangleq \sqrt{\eta}\hat{\mathbf{T}}_{\textrm{B}_{\textrm{r}}}^{\frac{1}{2}}\hat{\mathbf{H}}_{\textrm{G}}\hat{\mathbf{T}}_{\textrm{R}_{\textrm{t}}}^{\frac{1}{2}},
\end{eqnarray}
where $\hat{\mathbf{H}}_{\textrm{G}}\in\mathbb{C}^{M\times K}$ has i.i.d. $\mathcal{CN}(0,1)$ elements and
\begin{eqnarray}
\hat{\mathbf{T}}_{\textrm{B}_{\textrm{r}}}=\alpha_2^2\tau_2P_2 \eta\mathbf{T}_{\textrm{B}_{\textrm{r}}} \mathbf{R}_{\textrm{Z}_{\textrm{B}}}^{-1}\mathbf{T}_{\textrm{B}_{\textrm{r}}},
\end{eqnarray}
\vspace{-0.3cm}
\setlength{\arraycolsep}{-0.02em}
\begin{eqnarray}
\!\!\bar{\mathbf{T}}_{\textrm{R}_{\textrm{t}}}&&=\alpha_2^3\tau_2P_2 \eta^2 \biggl[\alpha_2\tau_2P_2\eta\textrm{tr}(\mathbf{T}_{\textrm{B}_{\textrm{r}}}\mathbf{R}_{\textrm{Z}_{\textrm{B}}}^{-1} \mathbf{T}_{\textrm{B}_{\textrm{r}}}^2\mathbf{R}_{\textrm{Z}_{\textrm{B}}}^{-1})\mathbf{T}_{\textrm{R}_{\textrm{t}}}\nonumber\\
&&\quad+K\left((1-\alpha_2)P_2\eta+\sigma_B^2\right) \textrm{tr}(\mathbf{R}_{\textrm{Z}_{\textrm{B}}}^{-1}\mathbf{T}_{\textrm{B}_{\textrm{r}}}^2\mathbf{R}_{\textrm{Z}_{\textrm{B}}}^{-1})\mathbf{I}_K\biggr].
\end{eqnarray}

Then the channel estimation error matrix is given by
\begin{eqnarray}
\tilde{\mathbf{G}}=\mathbf{G}- \hat{\mathbf{G}}\triangleq \sqrt{\eta}\tilde{\mathbf{T}}_{\textrm{B}_{\textrm{r}}}^{\frac{1}{2}}\tilde{\mathbf{H}}_{\textrm{G}}\tilde{\mathbf{T}}_{\textrm{R}_{\textrm{t}}}^{\frac{1}{2}},
\end{eqnarray}
where $\tilde{\mathbf{H}}_{\textrm{G}}\in\mathbb{C}^{M\times K}$ has i.i.d. $\mathcal{CN}(0,1)$ elements, $\tilde{\mathbf{T}}_{\textrm{B}_{\textrm{r}}}=\mathbf{T}_{\textrm{B}_{r}}-\hat{\mathbf{T}}_{\textrm{B}_{\textrm{r}}}$ and $\tilde{\mathbf{T}}_{\textrm{R}_{\textrm{t}}} =\frac{K}{\textrm{tr}\left(\check{\mathbf{T}}_{\textrm{R}_{\textrm{t}}}\right)}\check{\mathbf{T}}_{\textrm{R}_{\textrm{t}}}$ with $\check{\mathbf{T}}_{\textrm{R}_{\textrm{t}}}=M\eta \mathbf{T}_{\textrm{R}_{\textrm{t}}}-\bar{\mathbf{T}}_{\textrm{R}_{\textrm{t}}}$.

$\hat{\mathbf{T}}_{\textrm{R}_{\textrm{r}}}$, $\hat{\mathbf{T}}_{\textrm{R}_{\textrm{t}}}$, $\hat{\mathbf{T}}_{\textrm{B}_{\textrm{r}}}$, $\tilde{\mathbf{T}}_{\textrm{R}_{\textrm{r}}}$, $\tilde{\mathbf{T}}_{\textrm{R}_{\textrm{t}}}$, and $\tilde{\mathbf{T}}_{\textrm{B}_{\textrm{r}}}$ are positive semi-definite Hermitian matrices with $\hat{t}_{\textrm{R}_{\textrm{r}},ij}$, $\hat{t}_{\textrm{R}_{\textrm{t}},ij}$, $\hat{t}_{\textrm{B}_{\textrm{r}},ij}$, $\tilde{t}_{\textrm{R}_{\textrm{r}},ij}$, $\tilde{t}_{\textrm{R}_{\textrm{t}},ij}$, and $\tilde{t}_{\textrm{B}_{\textrm{r}},ij}$ denoting the corresponding $(i,j)$th elements, respectively. $\hat{\mathbf{D}}_{\textrm{F}}$ and $\tilde{\mathbf{D}}_{\textrm{F}}$ are diagonal matrices with $\hat{\beta}_i$ and $\tilde{\beta}_i$ representing the $i$th diagonal elements.

\vspace{-0.3cm}
\section{Achievable Rate Analysis}

In this section, we evaluate the achievable rate performance analytically with MRC processing and estimated CSI at the RS and the BS, based on which some useful insights are extracted for system deployment and design.

Since MRC processing and estimated CSI are used at the RS and the BS, we have $\mathbf{W}=\hat{\mathbf{F}}$ and $\mathbf{A}=\hat{\mathbf{G}}$. Then the $k$th element of $\mathbf{u}_{\textrm{B}}$ in (\ref{eqn_rB}) is given by
\setlength{\arraycolsep}{0.0em}
\begin{eqnarray}
\label{eqn_rBk}
\textrm{u}_{\textrm{B},k}&& =\underbrace{\alpha_{1}\alpha_{2}\kappa\sqrt{P_{\textrm{U}}}\hat{\mathbf{g}}_{k}^H\hat{\mathbf{G}}\hat{\mathbf{F}}^H\hat{\mathbf{f}}_{k}x_{k}}_{\textrm{ Desired signal}} \nonumber\\ &&\quad+\underbrace{\alpha_{1}\alpha_{2}\kappa\!\sqrt{P_{\textrm{U}}}\!\left(\!\hat{\mathbf{g}}_{k}^H\hat{\mathbf{G}}\hat{\mathbf{F}}^H\tilde{\mathbf{f}}_{k} \!+\!\hat{\mathbf{g}}_{k}^H\tilde{\mathbf{G}}\hat{\mathbf{F}}^H\hat{\mathbf{f}}_{k} \!+\!\hat{\mathbf{g}}_{k}^H\tilde {\mathbf{G}}\hat{\mathbf{F}}^H\tilde{\mathbf{f}}_{k}\!\right)\!x_{k}}_{\textrm{Interference-leakage due to channel estimation errors}} \nonumber\\
&&\quad+\underbrace{\alpha_{1}\alpha_{2}\kappa\sqrt{P_{\textrm{U}}}\sum_{j\neq k}^{K}\hat{\mathbf{g}}_{k}^H\mathbf{G}\hat{\mathbf{F}}^H\mathbf{f}_{j}x_{j}} _{\textrm{Inter-user interference}}\nonumber\\ &&\quad+\underbrace{\alpha_{1}\alpha_{2}\kappa\hat{\mathbf{g}}_{k}^H\mathbf{G}\hat{\mathbf{F}}^H\mathbf{n}_{\textrm{R}} +\alpha_{2}\kappa\hat{\mathbf{g}}_{k}^H\mathbf{G}\hat{\mathbf{F}}^H\mathbf{n}_{\textrm{q}1}} _{\textrm{AWGN and quantization noise at the RS}}\nonumber\\
&&\quad+\!\!\!\!\!\!\!\!\!\underbrace{\alpha_{2}\hat{\mathbf{g}}_{k}^H\mathbf{n}_{\textrm{B}}+\hat{\mathbf{g}}_{k}^H\mathbf{n}_{\textrm{q}2}}_{\textrm{AWGN and quantization noise at the BS}}\!\!\!\!\!\!\!\!\!.
\end{eqnarray}

Then the ergodic sum rate can be expressed as
\begin{eqnarray}
\label{eqn_Rsum}
R_{\textrm{sum}}=\mu\mathbb{E}\left\{\sum_{k=1}^K\log_2\left(1+\frac{\chi |\hat{\mathbf{g}}_{k}^H\hat{\mathbf{G}}\hat{\mathbf{F}}^H\hat{\mathbf{f}}_{k}|^2}{B_2}\right)\right\},
\end{eqnarray}
where $\mu=\frac{T-\tau_1-\tau_2}{2T}$, $\chi=\alpha_1^2\alpha_2^2\kappa^2P_{\textrm{U}}$, and $B_2=\chi(|B_1|^2+\sum_{j\neq k}^{K}|\hat{\mathbf{g}}_{k}^H\mathbf{G}\hat{\mathbf{F}}^H\mathbf{f}_{j}|^2) +\alpha_{1}^2\alpha_{2}^2\kappa^2\sigma_{\textrm{R}}^2\|\hat{\mathbf{g}}_{k}^H\mathbf{G}\hat{\mathbf{F}}^H\|^2 +\alpha_{2}^2\kappa^2|\hat{\mathbf{g}}_{k}^H\mathbf{G}\hat{\mathbf{F}}^H\mathbf{n}_{\textrm{q}1}|^2+\alpha_{2}^2\sigma_{\textrm{B}}^2\|\hat{\mathbf{g}}_{k}\|^2 +|\hat{\mathbf{g}}_{k}^H\mathbf{n}_{\textrm{q}2}|^2$ with $B_1=\hat{\mathbf{g}}_{k}^H\hat{\mathbf{G}}\hat{\mathbf{F}}^H\tilde{\mathbf{f}}_{k} +\hat{\mathbf{g}}_{k}^H\tilde{\mathbf{G}}\hat{\mathbf{F}}^H\hat{\mathbf{f}}_{k} +\hat{\mathbf{g}}_{k}^H\tilde {\mathbf{G}}\hat{\mathbf{F}}^H\tilde{\mathbf{f}}_{k}$. As shown by (\ref{eqn_Rsum}), evaluating the exact performance of the achievable rate requires the average over numerous channel realizations, which is quite time-consuming for the two-hop massive MIMO system due to the multiplication of large-dimensional matrices. Therefore, we derive a closed-form approximation of the achievable sum rate in the following theorem, proved in Appendix B.

\textbf{Theorem 1.} \emph{The achievable sum rate of $K$ users is approximated by}
\setlength{\arraycolsep}{0.0em}
\begin{eqnarray}
\label{eqn_app}
\hat{R}_{\textrm{sum}}=\mu\sum_{k=1}^K\log_2 \left(1+\frac{S_{k}}{I_{k}+N_{k1}+N_{k2}}\right),
\end{eqnarray}
\emph{where $S_{k}$, $I_{k1}$, $N_{k1}$, and $N_{k2}$ are given by (\ref{eqn_S_appk})$-$(\ref{eqn_N_k2}), shown at the top of the next page, with $\kappa$ given by Appendix B.1).}

\newcounter{mytempeqncnt1}
\begin{figure*}[!t]
\normalsize
\setcounter{mytempeqncnt1}{\value{equation}}
\setcounter{equation}{36}
\setlength{\arraycolsep}{-0.2em}
\begin{eqnarray}
S_{k}\!=\!\chi\eta^2\hat{t}_{\textrm{R}_{\textrm{t}},kk}^2\hat{\beta}_k^2\!\left(\!\textrm{tr}^2(\hat{\mathbf{T}}_{\textrm{B}_{\textrm{r}}})\! +\!\|\hat{\mathbf{T}}_{\textrm{B}_{\textrm{r}}}\|_{F}^2\!\right) \textrm{tr}^2(\hat{\mathbf{T}}_{\textrm{R}_{\textrm{r}}}) \!+\!\chi\hat{\beta}_k\|\hat{\mathbf{T}}_{\textrm{R}_{\textrm{r}}}\|_{F}^2\!\sum_{i=1}^K\hat{\beta}_i\!\left(\hat{t}_{\textrm{R}_{\textrm{t}},ki}^2\textrm{tr}^2(\hat{\mathbf{T}}_{\textrm{B}_{\textrm{r}}}) \!+\!\hat{t}_{\textrm{R}_{\textrm{t}},kk}\hat{t}_{\textrm{R}_{\textrm{t}},ii}\|\hat{\mathbf{T}}_{\textrm{B}_{\textrm{r}}}\|_{F}^2\right),\label{eqn_S_appk}
\end{eqnarray}
\vspace{-0.2cm}
\begin{eqnarray}
&&I_{k}=\chi\eta^2\sum_{j\neq k}^{K}\biggl[\hat{\beta}_j^2\left(\hat{t}_{\textrm{R}_{\textrm{t}},kj}^2\textrm{tr}^2(\hat{\mathbf{T}}_{\textrm{B}_{\textrm{r}}}) \!+\!\hat{t}_{\textrm{R}_{\textrm{t}},kk}\hat{t}_{\textrm{R}_{\textrm{t}},jj}\|\hat{\mathbf{T}}_{\textrm{B}_{\textrm{r}}}\|_{F}^2\right)\textrm{tr}^2(\hat{\mathbf{T}}_{\textrm{R}_{\textrm{r}}}) \!+\!\hat{\beta}_j\|\hat{\mathbf{T}}_{\textrm{R}_{\textrm{r}}}\|_{F}^2\!\sum_{i=1}^K\!\hat{\beta}_i\!\left(\hat{t}_{\textrm{R}_{\textrm{t}},ki}^2\textrm{tr}^2(\hat{\mathbf{T}}_{\textrm{B}_{\textrm{r}}}) \!+\!\hat{t}_{\textrm{R}_{\textrm{t}},kk}\hat{t}_{\textrm{R}_{\textrm{t}},ii}\|\hat{\mathbf{T}}_{\textrm{B}_{\textrm{r}}}\|_{F}^2\right)\biggr]\nonumber\\
&&\quad\quad+\chi\eta^2\sum_{i=1}^{K}\biggl[\tilde{\beta}_i \textrm{tr}(\tilde{\mathbf{T}}_{\textrm{R}_{\textrm{r}}}\hat{\mathbf{T}}_{\textrm{R}_{\textrm{r}}}) \sum_{l=1}^K\hat{\beta}_l\left(\hat{t}_{\textrm{R}_{\textrm{t}},kl}^2\textrm{tr}^2(\hat{\mathbf{T}}_{\textrm{B}_{\textrm{r}}}) +\hat{t}_{\textrm{R}_{\textrm{t}},kk}\hat{t}_{\textrm{R}_{\textrm{t}},ll}\|\hat{\mathbf{T}}_{\textrm{B}_{\textrm{r}}}\|_{F}^2\right) +\hat{t}_{\textrm{R}_{\textrm{t}},kk}\tilde{t}_{\textrm{R}_{\textrm{t}},ii}\hat{\beta}_i^2\textrm{tr}(\hat{\mathbf{T}}_{\textrm{B}_{\textrm{r}}}\tilde{\mathbf{T}}_{\textrm{B}_{\textrm{r}}}) \textrm{tr}^2(\hat{\mathbf{T}}_{\textrm{R}_{\textrm{r}}}) \nonumber\\
&&\quad\quad\quad+\hat{\beta}_i\sum_{l=1}^K \hat{t}_{\textrm{R}_{\textrm{t}},kk}\tilde{t}_{\textrm{R}_{\textrm{t}},ll}\hat{\beta}_l\textrm{tr}(\hat{\mathbf{T}}_{\textrm{B}_{\textrm{r}}}\tilde{\mathbf{T}}_{\textrm{B}_{\textrm{r}}}) \|\hat{\mathbf{T}}_{\textrm{R}_{\textrm{r}}}\|_{F}^2 +\hat{t}_{\textrm{R}_{\textrm{t}},kk}\tilde{\beta}_i\textrm{tr}(\hat{\mathbf{T}}_{\textrm{B}_{\textrm{r}}}\tilde{\mathbf{T}}_{\textrm{B}_{\textrm{r}}}) \textrm{tr}(\hat{\mathbf{T}}_{\textrm{R}_{\textrm{r}}}\tilde{\mathbf{T}}_{\textrm{R}_{\textrm{r}}}) \sum_{l=1}^K\tilde{t}_{\textrm{R}_{\textrm{t}},ll} \hat{\beta}_l\biggr]\label{eqn_I_k2}
\end{eqnarray}
\vspace{-0.2cm}
\begin{eqnarray}
&&N_{k1}=\alpha_1(1-\alpha_1)\alpha_2^2\kappa^2P_{\textrm{U}}\eta^2\biggl[\sum_{n=1}^{N}\hat{t}_{\textrm{R}_{\textrm{r}},nn}^2\sum_{i=1}^K\hat{\beta}_i\left(\hat{t}_{\textrm{R}_{\textrm{t}},ki}^2\textrm{tr}^2(\hat{\mathbf{T}}_{\textrm{B}_{\textrm{r}}}) +\hat{t}_{\textrm{R}_{\textrm{t}},kk}\left(\hat{t}_{\textrm{R}_{\textrm{t}},ii}\|\hat{\mathbf{T}}_{\textrm{B}_{\textrm{r}}}\|_{F}^2 +\tilde{t}_{\textrm{R}_{\textrm{t}},ii}\textrm{tr}(\hat{\mathbf{T}}_{\textrm{B}_{\textrm{r}}}\tilde{\mathbf{T}}_{\textrm{B}_{\textrm{r}}})\right)\right)\left( \hat{\beta}_i+\sum_{l=1}^K\hat{\beta}_l\right)\nonumber\\
&&\quad\quad+\sum_{n=1}^{N}\hat{t}_{\textrm{R}_{r},nn}\tilde{t}_{\textrm{R}_{\textrm{r}},nn}\sum_{i=1}^K\hat{\beta}_i \left(\hat{t}_{\textrm{R}_{\textrm{t}},ki}^2\textrm{tr}^2(\hat{\mathbf{T}}_{\textrm{B}_{\textrm{r}}}) +\hat{t}_{\textrm{R}_{\textrm{t}},kk}\left(\hat{t}_{\textrm{R}_{\textrm{t}},ii}\|\hat{\mathbf{T}}_{\textrm{B}_{\textrm{r}}}\|_{F}^2 +\tilde{t}_{\textrm{R}_{\textrm{t}},ii}\textrm{tr}(\hat{\mathbf{T}}_{\textrm{B}_{\textrm{r}}}\tilde{\mathbf{T}}_{\textrm{B}_{\textrm{r}}})\right)\right)\sum_{l=1}^K\tilde{\beta}_l\biggr]\nonumber\\
&&\quad\quad\quad+\alpha_1\alpha_2^2\kappa^2\eta^2\sigma_{\textrm{R}}^2\biggl[\textrm{tr}(\hat{\mathbf{T}}_{\textrm{R}_{\textrm{r}}}) \sum_{i=1}^K\hat{\beta}_i\left(\hat{t}_{\textrm{R}_{\textrm{t}},ki}^2\textrm{tr}^2(\hat{\mathbf{T}}_{\textrm{B}_{\textrm{r}}}) +\hat{t}_{\textrm{R}_{\textrm{t}},kk}\hat{t}_{\textrm{R}_{\textrm{t}},ii}\|\hat{\mathbf{T}}_{\textrm{B}_{\textrm{r}}}\|_{F}^2\right) +\hat{t}_{\textrm{R}_{\textrm{t}},kk}\textrm{tr}(\hat{\mathbf{T}}_{\textrm{R}_{\textrm{r}}}) \textrm{tr}(\hat{\mathbf{T}}_{\textrm{B}_{\textrm{r}}}\tilde{\mathbf{T}}_{\textrm{B}_{\textrm{r}}})\sum_{i=1}^K\hat{\beta}_i \tilde{t}_{\textrm{R}_{\textrm{t}},ii}\biggr]\label{eqn_N_k1}
\end{eqnarray}
\vspace{-0.2cm}
\begin{eqnarray}
&&N_{k2}\!=\!\frac{\alpha_2(1\!-\!\alpha_2)P_{\textrm{R}}\eta^2}{K}\biggl[\sum_{m=1}^M \hat{t}_{\textrm{B}_{\textrm{\textrm{r}}},mm}^2\sum_{i=1}^K\left(\hat{t}_{\textrm{R}_{\textrm{t}},ki}^2\!+\!\hat{t}_{\textrm{R}_{\textrm{t}},kk}\hat{t}_{\textrm{R}_{\textrm{t}},ii}\right) \!+\!\hat{t}_{\textrm{R}_{\textrm{t}},kk}\sum_{m=1}^M \hat{t}_{\textrm{B}_{\textrm{r}},mm}\tilde{t}_{\textrm{B}_{\textrm{r}},mm}\sum_{i=1}^K\tilde{t}_{\textrm{R}_{\textrm{t}},ii}\biggr] \!+\!\alpha_2\eta\sigma_{\textrm{B}}^2\hat{t}_{\textrm{R}_{\textrm{t}},kk}\textrm{tr}(\hat{\mathbf{T}}_{\textrm{B}_{\textrm{r}}}).\label{eqn_N_k2}
\end{eqnarray}
\setcounter{equation}{\value{mytempeqncnt1}}
\hrulefill
\vspace*{-3pt}
\end{figure*}

For (\ref{eqn_app}), $S_{k}$, $I_{k}$, $N_{k1}$, and $N_{k2}$ denote the average powers of the desired signal, the aggregated interference caused by channel estimation errors and other users, the aggregated noise including quantization errors and AWGN at the RS, and the aggregated noise including quantization errors and AWGN at the BS, respectively. The theoretical approximation given by Theorem 1 addresses the time-consuming problem when using (\ref{eqn_Rsum}) and makes the performance evaluation flexible. It can be also seen that the achievable rate increases with the ADC resolutions at the RS and the BS.

Based on Theorem 1, we can further extract some insights to well dissect the considered system. Without loss of generality, we assume $M=\delta N$ with $\delta$ bounded, i.e., $\delta<\infty$, hereinafter.

\textbf{Proposition 1.} \emph{If perfect CSI is available at both the RS and the BS, we have $\hat{\mathbf{T}}_{\textrm{R}_{\textrm{r}}}=\mathbf{T}_{\textrm{R}_{\textrm{r}}}$, $\hat{\mathbf{D}}_{\textrm{F}}=\mathbf{D}_{\textrm{F}}$, $\hat{\mathbf{T}}_{\textrm{B}_{\textrm{r}}}=\mathbf{T}_{\textrm{B}_{\textrm{r}}}$, $\hat{\mathbf{T}}_{\textrm{R}_{\textrm{t}}}=\mathbf{T}_{\textrm{R}_{\textrm{t}}}$, and $\tilde{\mathbf{T}}_{\textrm{R}_{\textrm{r}}} =\tilde{\mathbf{D}}_{\textrm{F}}=\tilde{\mathbf{T}}_{\textrm{B}_{\textrm{r}}}=\tilde{\mathbf{T}}_{\textrm{R}_{\textrm{t}}}=\mathbf{0}$, which leads to}
\setcounter{equation}{40}
\begin{eqnarray}
\label{eqn_Sk_pCSI}
S_{k}&&=\chi\eta^2\beta_k^2\left(M^2\!+\!\|\mathbf{T}_{\textrm{B}_{\textrm{r}}}\|_{F}^2\right)N^2 \nonumber\\ &&+\chi\eta^2\beta_k\|\mathbf{T}_{\textrm{R}_{\textrm{r}}}\|_{F}^2\sum_{i=1}^K\beta_i\!\left(t_{\textrm{R}_{\textrm{t}},ki}^2M^2\!+\!\|\mathbf{T}_{\textrm{B}_{\textrm{r}}}\|_{F}^2\right),
\end{eqnarray}
\vspace{-0.3cm}
\begin{eqnarray}
\label{eqn_Ik_pCSI}
I_{k}&&=\chi\eta^2\sum_{j\neq k}^{K}\biggl[\beta_j^2\left(t_{\textrm{R}_{\textrm{t}},kj}^2M^2 \!+\!\|\mathbf{T}_{\textrm{B}_{\textrm{r}}}\|_{F}^2\right)N^2 \nonumber\\
&&+\beta_j\|\mathbf{T}_{\textrm{R}_{\textrm{r}}}\|_{F}^2\!\sum_{i=1}^K\!\beta_i\!\left(t_{\textrm{R}_{\textrm{t}},ki}^2M^2 \!+\!\|\mathbf{T}_{\textrm{B}_{\textrm{r}}}\|_{F}^2\right)\biggr],
\end{eqnarray}
\vspace{-0.3cm}
\begin{eqnarray}
\label{eqn_Nk1_pCSI}
\!\!N_{k1}&&=\alpha_1(1-\alpha_1)\alpha_2^2\kappa^2P_{\textrm{U}}\eta^2 \nonumber\\
&&\times\sum_{i=1}^K\beta_i\!\biggl(\!\beta_i\!+\!\sum_{l=1}^K\beta_l\!\biggr)\!\!\left(t_{\textrm{R}_{\textrm{t}},ki}^2 M^2\!\!+\!\|\mathbf{T}_{\textrm{B}_{\textrm{r}}}\|_{F}^2\right)\!N \nonumber\\
&&+\alpha_1\alpha_2^2\kappa^2\eta^2\sigma_{\textrm{R}}^2\sum_{i=1}^K\beta_i (t_{\textrm{R}_{\textrm{t}},ki}^2M^2 +\|\mathbf{T}_{\textrm{B}_{\textrm{r}}}\|_{F}^2)N,
\end{eqnarray}
\vspace{-0.3cm}
\begin{eqnarray}
\label{eqn_Nk2_pCSI}
N_{k2}&&=\frac{\alpha_2(1\!-\!\alpha_2)P_{\textrm{R}}\eta^2}{K}\!\biggl(\!1\!+\!\frac{1}{K}\sum_{i=1}^K t_{\textrm{R}_{\textrm{t}},ki}^2\!\biggr)\!M \!+\!\alpha_2\eta\sigma_{\textrm{B}}^2 M,\quad
\end{eqnarray}
\emph{where $t_{\textrm{R}_{\textrm{t}},ij}$ denotes the $(i,j)$th element of $\mathbf{T}_{\textrm{R}_{\textrm{t}}}$.}

For $\mathbf{T}_{\textrm{B}_{\textrm{r}}}$, we have
\begin{eqnarray}
\|\mathbf{T}_{\textrm{B}_{\textrm{r}}}\|_{F}^2&&=\textrm{tr}(\mathbf{T}_{\textrm{B}_{\textrm{r}}}^H\mathbf{T}_{\textrm{B}_{\textrm{r}}}) =\sum_{m=1}^M \lambda_{\textrm{B}_{\textrm{r}},m}^2 \leq \lambda_{\textrm{B}_{\textrm{r}},1}\sum_{m=1}^M \lambda_{\textrm{B}_{\textrm{r}},m}\nonumber\\
&&=\rho(\mathbf{T}_{\textrm{B}_{\textrm{r}}})\textrm{tr}(\mathbf{T}_{\textrm{B}_{\textrm{r}}})=\|\mathbf{T}_{\textrm{B}_{\textrm{r}}}\|_2M,
\end{eqnarray}
where $\lambda_{\textrm{B}_{\textrm{r}},m}$ and $\rho(\mathbf{T}_{\textrm{B}_{\textrm{r}}})$ denote the $m$th largest eigenvalue and spectral radius of $\mathbf{T}_{\textrm{B}_{\textrm{r}}}$, respectively. Since $\|\mathbf{T}_{\textrm{B}_{\textrm{r}}}\|_2$ is uniformly bounded and $M=\delta N$, we have $\|\mathbf{T}_{\textrm{B}_{\textrm{r}}}\|_{F}^2=\mathcal{O}(N)$. Similarly, we can obtain $\|\mathbf{T}_{\textrm{R}_{\textrm{r}}}\|_{F}^2= \mathcal{O}(N)$. Therefore, the achievable rate is able to grow unboundedly with $N$ even in presence of spatial correlation at the RS and the BS. In addition, the results in Proposition 1 can be directly reduced to \cite[Theorem 1]{P. Dong_a} provided that no channel correlation exists for both two hops and ideal ADCs are used at the BS. Based on Proposition 1, the following corollaries demonstrate the further impacts of power scaling, channel correlation, and ADC resolution on the achievable rate performance.

\textbf{Corollary 1.} \emph{(Power Scaling Law) Let $\gamma_{k}=\frac{S_{k}}{I_{k}+N_{k1}+N_{k2}}$ and scale down the transmit powers as $P_{\textrm{U}}=\frac{E_{\textrm{U}}}{N^a}$ and $P_{\textrm{R}}=\frac{E_{\textrm{R}}}{M^b}$ with $E_{\textrm{U}}$, $E_{\textrm{R}}$ fixed and $a,b \geq 0$. When $N$ grows to infinity, we have}
\begin{eqnarray}
\label{eqn_lim_gammak}
\lim\limits_{N \to \infty}\gamma_{k}=
\begin{cases}
\infty, & a,b<1\\
\frac{\alpha_2\beta_k^2 \eta E_{\textrm{R}}}{\sigma_{\textrm{B}}^2 \sum_{i=1}^K \beta_i^2}, & a<b=1\\
\frac{\alpha_1 \beta_k E_{\textrm{U}}}{\sigma_{\textrm{R}}^2}, & b<a=1\\
\frac{\alpha_1\alpha_2 \beta_k^2\eta E_{\textrm{U}} E_{\textrm{R}}}{\zeta}, & a=b=1\\
0, & a>1 \ \text{or} \ b>1
\end{cases},
\end{eqnarray}
where $\zeta=\alpha_2\beta_k\eta\sigma_{\textrm{R}}^2 E_{\textrm{R}}+\sigma_{\textrm{B}}^2\left(\alpha_1 E_{\textrm{U}}\sum_{i=1}^K \beta_i^2 +\sigma_{\textrm{R}}^2\sum_{i=1}^K \beta_i\right)$.
\begin{IEEEproof}
Please see Appendix C.
\end{IEEEproof}

Corollary 1 reveals that the impacts of ADC resolution at the RS and the BS vanish asymptotically when the power is scaled as $a<b=1$ and $b<a=1$, respectively, while both of them persist with $a=b=1$. By letting $\alpha_2=1$, Corollary 1 coincides with \cite[Proposition 1]{P. Dong_a}, which demonstrates that the channel correlation does not impact the asymptotic behavior of the achievable rate with power scaling. In the imperfect CSI case, the power scaling law is still valid to guarantee the non-vanishing asymptotic rate so long as the transmit powers for channel estimation are fixed. Note that the transmit powers for channel estimation are usually fixed instead of scaled since accurate channel estimates are important for the following signal detection and decoding. Different from the perfect CSI case, the ADC resolution at the RS will also impact the bounded asymptotic rate with $a<b=1$ in addition to the impact of low-resolution ADCs at the BS for imperfect CSI case.

\textbf{Corollary 2.} \emph{(Impact of RS Transmit Correlation) By selecting $K$ equally spaced antennas from $N$ RS antennas for signal forwarding, the impact of RS transmit correlation vanishes as $N\rightarrow\infty$.}
\begin{IEEEproof}
It can be easily proved according to (\ref{eqn_lim_tRt_ij2}) and (\ref{eqn_lim_tRt_ijM2}).
\end{IEEEproof}

\textbf{Corollary 3.} \emph{(Impacts of RS and BS Receive Correlation) As $N\rightarrow\infty$, the impacts of receive correlation at the RS and the BS are dependent on $\delta$, i.e.}
\begin{eqnarray}
\begin{cases}
\lim\limits_{N \to \infty}\hat{R}_{\textrm{sum}}^{\mathbf{T}_{\textrm{R}_{\textrm{r}}}=\mathbf{I}_{N}}< \lim\limits_{N \to \infty}\hat{R}_{\textrm{sum}}^{\mathbf{T}_{\textrm{B}_{\textrm{r}}}=\mathbf{I}_{M}}, & \delta<1\\
\lim\limits_{N \to \infty}\hat{R}_{\textrm{sum}}^{\mathbf{T}_{\textrm{R}_{\textrm{r}}}=\mathbf{I}_{N}}\approx \lim\limits_{N \to \infty}\hat{R}_{\textrm{sum}}^{\mathbf{T}_{\textrm{B}_{\textrm{r}}}=\mathbf{I}_{M}}, & \delta=1\\
\lim\limits_{N \to \infty}\hat{R}_{\textrm{sum}}^{\mathbf{T}_{\textrm{R}_{\textrm{r}}}=\mathbf{I}_{N}}> \lim\limits_{N \to \infty}\hat{R}_{\textrm{sum}}^{\mathbf{T}_{\textrm{B}_{\textrm{r}}}=\mathbf{I}_{M}}, & \delta>1
\end{cases}.
\end{eqnarray}
\begin{IEEEproof}
Please see Appendix D.
\end{IEEEproof}

Corollary 3 reveals that the relationship between the numbers of antennas at the RS and the BS determines which receive correlation incurs more rate loss. If the number of antennas at the RS is larger than that at the BS, i.e., $\delta<1$, a rich scattering environment around the BS will be very helpful to improve the performance while the environment around the RS is relatively trivial. If the RS is equipped with fewer antennas than the BS, i.e., $\delta>1$, weakening the receive correlation at the RS can take back more rate performance, which indicates that deploying the RS at the rich scattering environment is recommended. On the other hand, Corollary 3 also indicates how many antennas should be equipped at the RS and the BS according to the scattering environments around them. The BS should be usually deployed on the top of buildings while the RS can be deployed flexibly, which makes the rich scattering environment relatively easy to achieve for the RS. Then we can equip more antennas at the BS compared to the RS and deploy the RS at a rich scattering environment to significantly improve the performance.


\textbf{Corollary 4.} \emph{(Impacts of ADC resolution at RS and BS) As $N\rightarrow\infty$, the impacts of ADC resolution at the RS and the BS are dependent on $\delta$, i.e.}
\begin{eqnarray}
\begin{cases}
\lim\limits_{N \to \infty}\hat{R}_{\textrm{sum}}^{\alpha_1=1}< \lim\limits_{N \to \infty}\hat{R}_{\textrm{sum}}^{\alpha_2=1}, & \delta<1\\
\lim\limits_{N \to \infty}\hat{R}_{\textrm{sum}}^{\alpha_1=1}\approx \lim\limits_{N \to \infty}\hat{R}_{\textrm{sum}}^{\alpha_2=1}, & \delta=1\\
\lim\limits_{N \to \infty}\hat{R}_{\textrm{sum}}^{\alpha_1=1}> \lim\limits_{N \to \infty}\hat{R}_{\textrm{sum}}^{\alpha_2=1}, & \delta>1
\end{cases}.
\end{eqnarray}
\begin{IEEEproof}
Please see Appendix E.
\end{IEEEproof}

From Corollary 4, the relationship between the numbers of antennas at the RS and the BS also indicates how many ADC resolution bits the RS and the BS should adopt to achieve better performance. If the RS is equipped with more antennas than the BS, i.e., $\delta<1$, it is better to use higher resolution ADCs at the BS while the requirement on the ADC resolution at the RS will be relatively relaxed. On the contrary, if the number of antennas at the BS is larger than that at the RS, i.e., $\delta>1$, the requirement on ADC resolution is transferred to the RS and using higher resolution ADCs is preferable. For the BS and the RS, the one equipped with fewer antennas should use higher resolution\footnote{The ``higher resolution" in this paper refers to relatively high resolution for the BS/RS compared to the RS/BS, e.g., three-bit vs. one-bit, but is not necessarily the very high resolution larger than eight bits.} ADCs while another one with more antennas can safely use lower resolution ADCs since the antennas can provide additional array gains to compensate the performance loss.


To sum up, the relationship between the numbers of antennas at the RS and the BS is the key parameter that determines how the spatial correlation and ADC resolution impact the performance. The requirement on the spatial correlation and ADC resolution is relatively relaxed if one can be equipped with more antennas than another one for the RS and the BS. If the favorable scattering environment is difficult to achieve for the BS, we can accordingly adjust the numbers of antennas at the RS and the BS and the ADC resolution bits to guarantee the satisfactory performance.

\section{Simulation Results}

In this section, numerical results are presented to verify our theoretical analysis and to show the impacts of the key parameters on the achievable rate performance straightforwardly. For the two-hop channels, the large-scale fading is modeled as $\beta_k=(d_{\textrm{ref}}/d_{\textrm{U}_{k}\textrm{R}})^\nu$ and $\eta=(d_{\textrm{ref}}/d_{\textrm{RB}})^\nu$, respectively \cite{P. Dong_d}, where $d_{\textrm{ref}}$, $d_{\textrm{U}_{k}\textrm{R}}$, $d_{\textrm{RB}}$, and $\nu$ denote the reference distance, the distance between the $k$th user and the RS, the distance between the RS and the BS, and the path loss exponent, respectively. The representative and widely used exponential correlation model is used to depict the spatial correlation at the RS and the BS, that is
\begin{eqnarray}
\label{correlation_matrix}
[\mathbf{T}_{\star}]_{ij}=
\begin{cases}
r_{\star}^{j-i}, & i\leq j,\\
(r_{\star}^{i-j})^{*}, & i>j,
\end{cases}
\end{eqnarray}
where $\star\in\{\textrm{R}_{\textrm{r}}, \textrm{R}_{\textrm{t}}, \textrm{B}_{\textrm{r}}\}$. By selecting $K$ equally spaced antennas for signal retransmission at the RS, we have $r_{\textrm{R}_{\textrm{t}}}=r_{\textrm{R}_{\textrm{r}}}^{\frac{N}{K}} =r_{\textrm{R}}^{\frac{N}{K}}$ with $r_{\textrm{R}}$ denoting the correlation coefficient of adjacent antennas at the RS. In addition, $r_{\textrm{B}_{\textrm{r}}}=r_{\textrm{B}}$  denotes the correlation coefficient of adjacent antennas at the BS. Unless stated otherwise, the parameter settings are detailed in Table~\ref{Sim_para}.

\begin{table}
  \centering
  \caption{Setting Values of Simulation Parameters}\label{Sim_para}
  \begin{tabular}{|c|c|}
  \hline
  Simulation Parameter & Setting Value\\
  \hline
  $T$ & 100\\
  \hline
  $K$ & 10\\
  \hline
  $\delta$ & 2\\
  \hline
  $E_\textrm{U}$, $P_\textrm{1}$ & 20 dB\\
  \hline
  $E_\textrm{R}$, $P_\textrm{2}$ & 25 dB\\
  \hline
  $\sigma_{\textrm{B}}^2$ & 1.5 dB\\
  \hline
  $\sigma_{\textrm{R}}^2$ & 2.2 dB\\
  \hline
  $d_{\text{ref}}$ & 100 meters\\
  \hline
  $d_{\textrm{U}_{k}\textrm{R}}$& \makecell{[182\   209\  197\   214\   190\   188\   \\
  201\   215\   206\   216] meters}\\
  \hline
  $d_{\textrm{RB}}$ & 250 meters\\
  \hline
  $\nu$ & 3.8\\
  \hline
  $r_{\textrm{R}}$, $r_{\textrm{B}}$ & 0.8\\
  \hline
  $a$, $b$ & 0\\
  \hline
  \end{tabular}
\end{table}

\begin{figure}[!t]
\centering
\includegraphics[trim=0 0 0 0, width=3.6in]{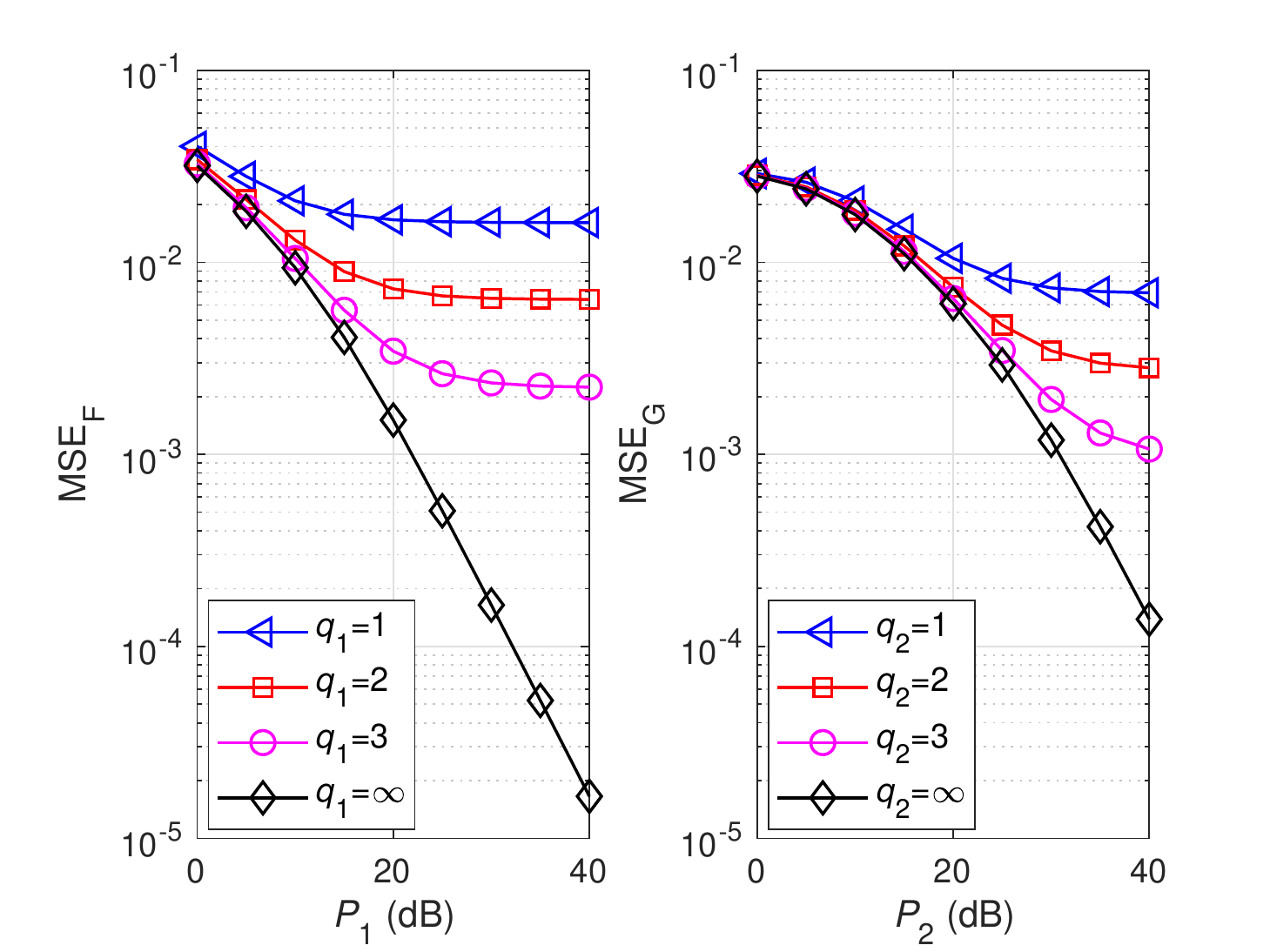}
\caption{Channel estimation performance of $\mathbf{F}$ and $\mathbf{G}$.}\label{MSE}
\end{figure}

In Fig.~\ref{MSE}, the channel estimation performance is first evaluated with different ADC resolution bits at the RS and the BS. The per element MSEs for $\mathbf{F}$ and $\mathbf{G}$ are defined as $\textrm{MSE}_{\textrm{F}}=\frac{\mathbb{E}\{\|\hat{\mathbf{F}}-\mathbf{F}\|^2_F\}}{NK}$ and $\textrm{MSE}_{\textrm{G}}=\frac{\mathbb{E}\{\|\hat{\mathbf{G}}-\mathbf{G}\|^2_F\}}{MK}$, respectively. For each hop, the estimation error decreases as the ADC resolution at the destination (RS or BS) becomes high. If using low-resolution ADCs, the MSE performance is limited by the obvious error floor as the transmit power grows to infinity.

\begin{figure}[!t]
\centering
\includegraphics[trim=0 0 0 0, width=3.6in]{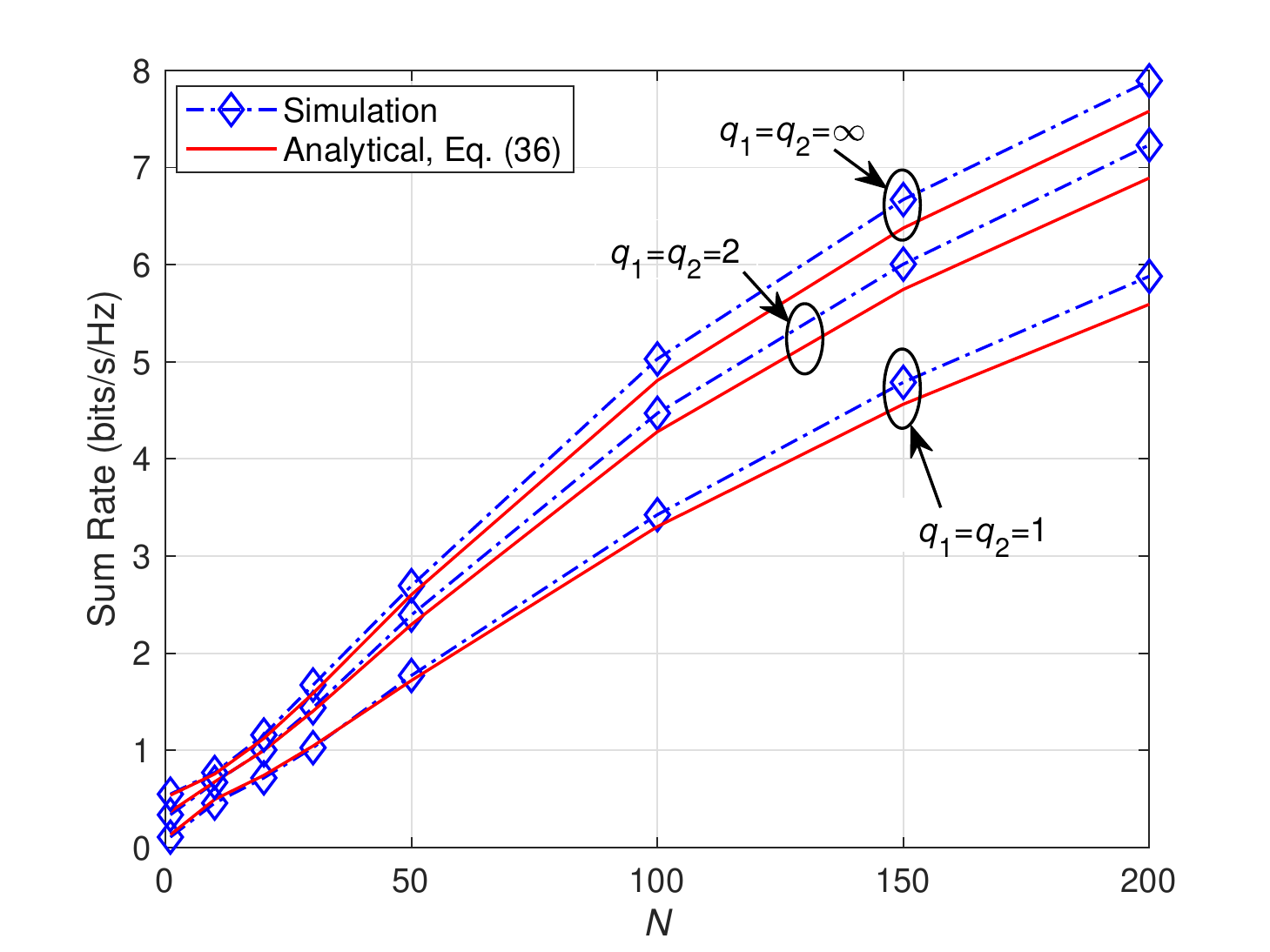}
\caption{Comparison between simulation and analytical results for the achievable rate with different ADC resolution bits.}\label{sim_theo_q}
\end{figure}

Fig.~\ref{sim_theo_q} compares the simulation results and the analytical results given by (\ref{eqn_app}) for the achievable rate with $q_1=q_2=1,2,\infty$. From this figure, the derived analytical expression faithfully reflect the simulated results and thus can be safely used to replace the time-consuming averaging over numerous channel realizations when evaluating the performance. Using the analytical expression becomes more effective as the number of antennas increases. In addition, if both the RS and the BS employ the one-bit ADCs, there is a significant rate gap compared to the case with ideal ADCs. When the resolution bits increase to two, the performance becomes fairly good with the very low cost.

\begin{figure}[!t]
\centering
\includegraphics[trim=0 0 0 0, width=3.6in]{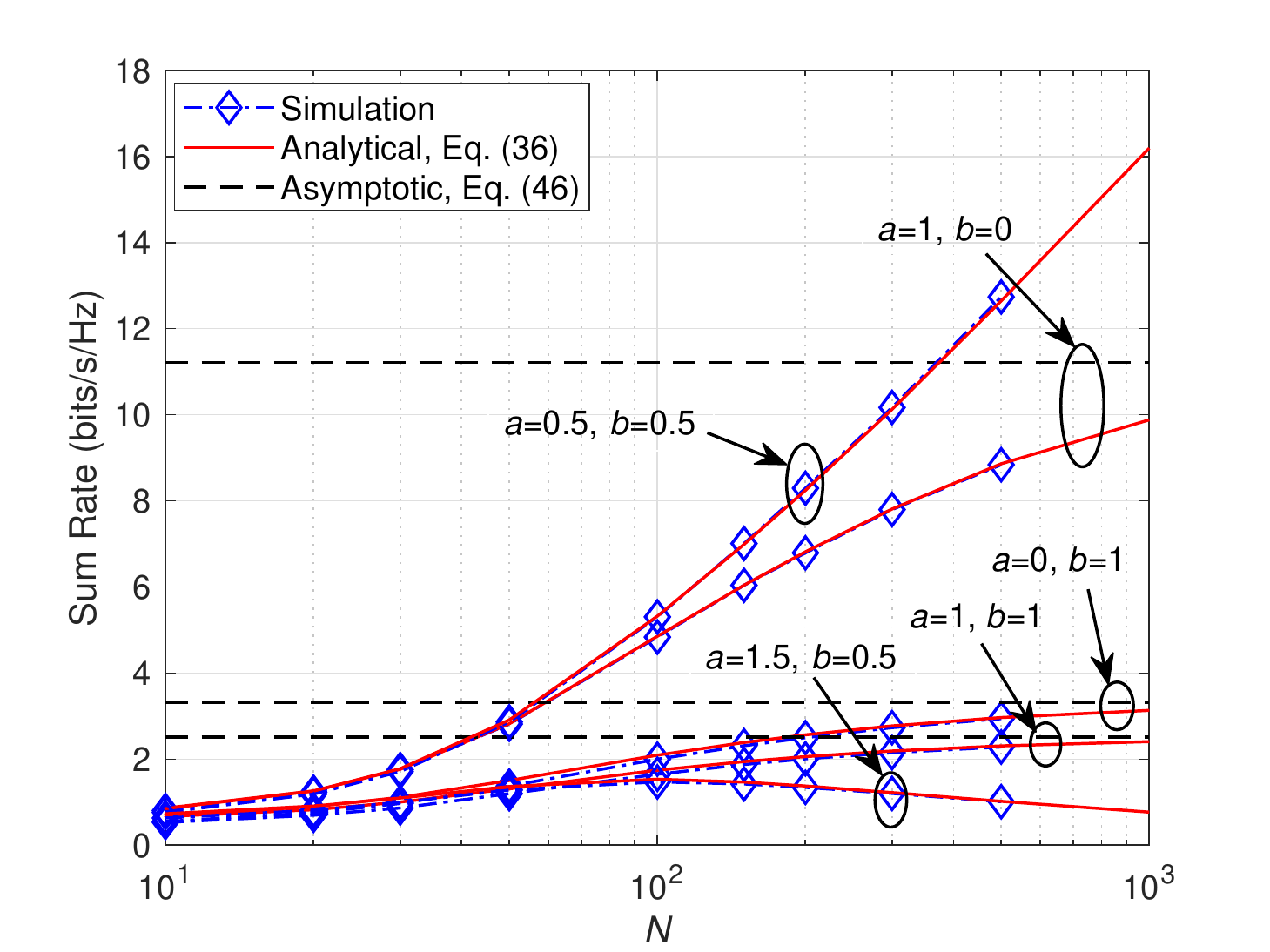}
\caption{Power scaling law with $q_1=q_2=2$ and perfect CSI.}\label{p_power_scaling}
\end{figure}

\begin{figure}[!t]
\centering
\includegraphics[trim=0 0 0 0, width=3.6in]{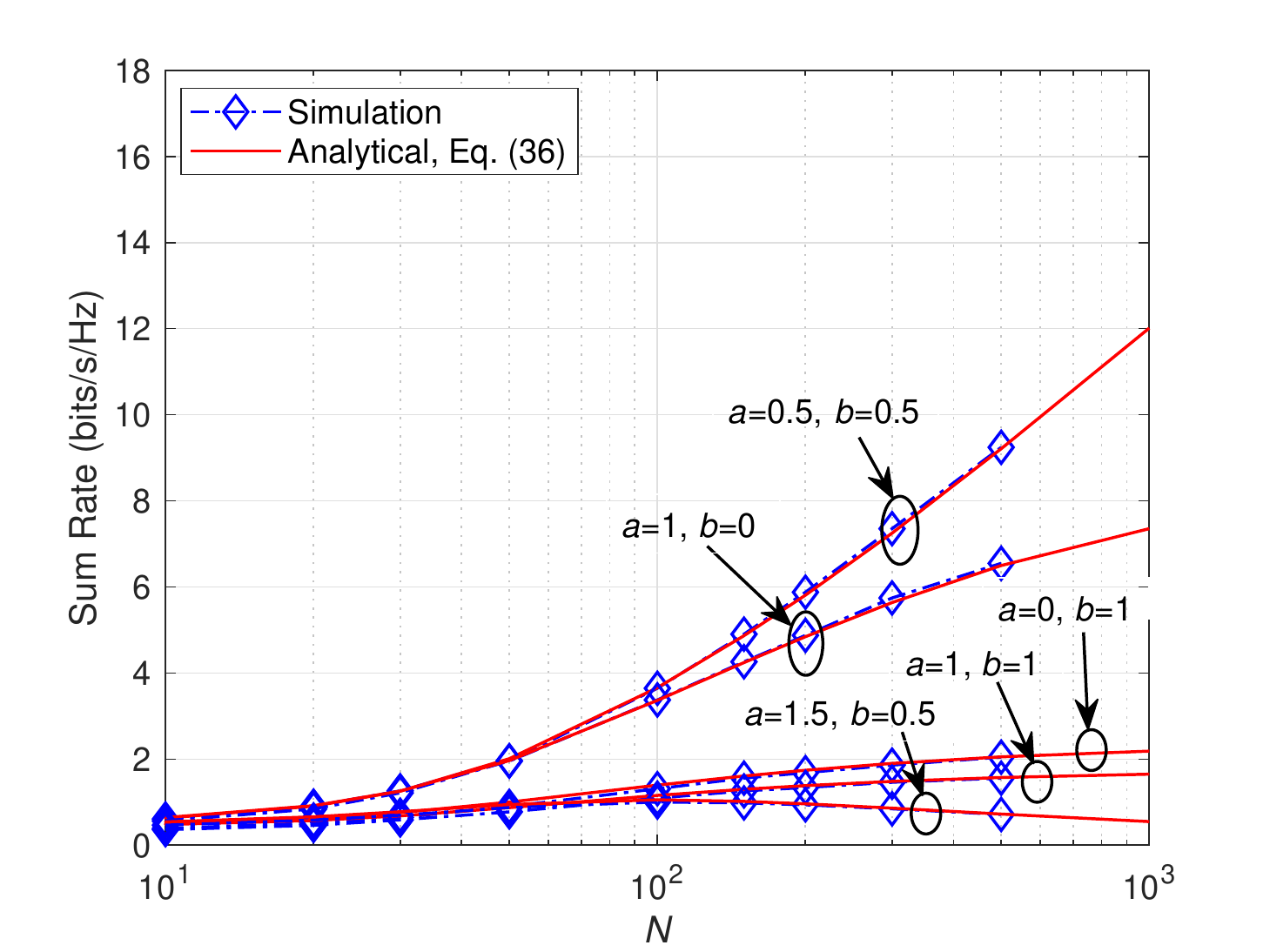}
\caption{Power scaling law with $q_1=q_2=2$ and imperfect CSI.}\label{ip_power_scaling}
\end{figure}

Figs.~\ref{p_power_scaling} and \ref{ip_power_scaling} show the power scaling laws of the considered system with perfect and imperfect CSI, respectively. The ADC resolution bits at the RS and the BS are fixed as $q_1=q_2=2$. Both the closed-form approximation and asymptotic limits respectively given by (\ref{eqn_app}) and (\ref{eqn_lim_gammak}) are verified. By comparing Figs.~\ref{p_power_scaling} and \ref{ip_power_scaling} with Figs.~3 and 4 in \cite{P. Dong_a}, it clearly demonstrates that the spatial correlation does not impact the asymptotic behavior of the achievable rate with power scaling, as discussed after Corollary 1. For the imperfect CSI case, by fixing transmit powers for channel estimation, users and the RS can scale their transmit powers for data transmission similarly to the perfect CSI case.

\begin{figure}[!t]
\centering
\includegraphics[trim=0 0 0 0, width=3.6in]{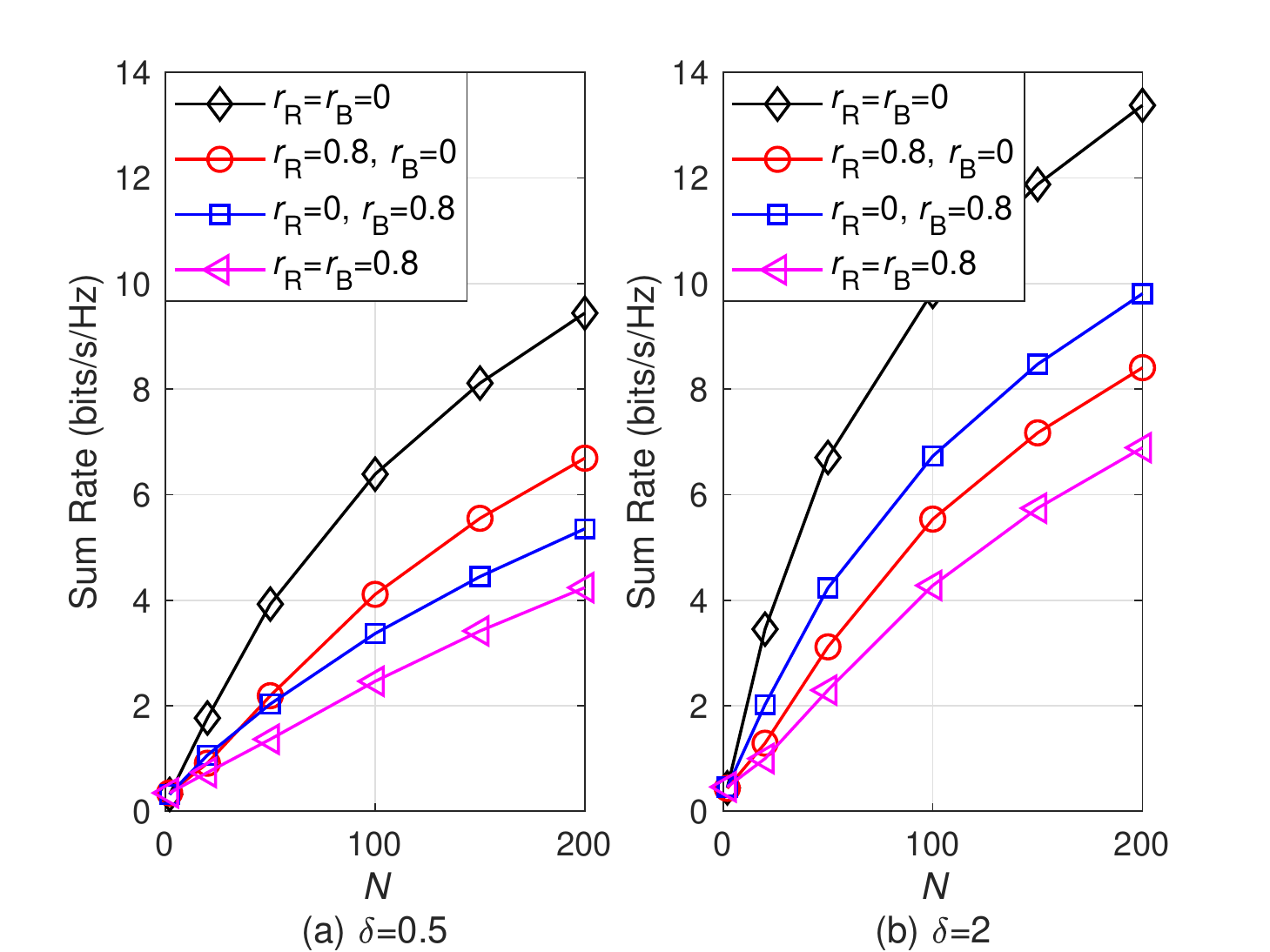}
\caption{Achievable rate performance with $q_1=q_2=2$ and different spatial correlation levels.}\label{impact_correlation}
\end{figure}

Fig.~\ref{impact_correlation} plots the achievable rate performance with $q_1=q_2=2$ and different spatial correlation levels to reveal the impacts of spatial correlation at the RS and the BS. Both the cases with $\delta=0.5$ and $\delta=2$ are considered to verify Corollary 3 and demonstrate the insights for system design straightforwardly. It is clear that the case with $r_{\textrm{R}}=0.8, r_{\textrm{B}}=0$ (rich scattering environment for the BS) and the case with $r_{\textrm{R}}=0, r_{\textrm{B}}=0.8$ (rich scattering environment for the RS) respectively achieve better performance in $\delta=0.5$ and $\delta=2$. If the scattering environment for the BS is poor, the performance can be improved remarkably by equipping more antennas at the BS compared to the RS and deploying the RS at a rich scattering environment. For the case with $r_{\textrm{R}}=0.8, r_{\textrm{B}}=0$, the performance improvement is not so significant through the same way. For example, when $N=200$, the sum rate is improved from $5.4$ bits/s/Hz in $\delta=0.5$ to $9.8$ bits/s/Hz in $\delta=2$ for $r_{\textrm{R}}=0, r_{\textrm{B}}=0.8$ while the improvement for $r_{\textrm{R}}=0.8, r_{\textrm{B}}=0$ is merely from $6.7$ bits/s/Hz to $8.4$ bits/s/Hz.


\begin{figure}[!t]
\centering
\includegraphics[trim=0 0 0 0, width=3.6in]{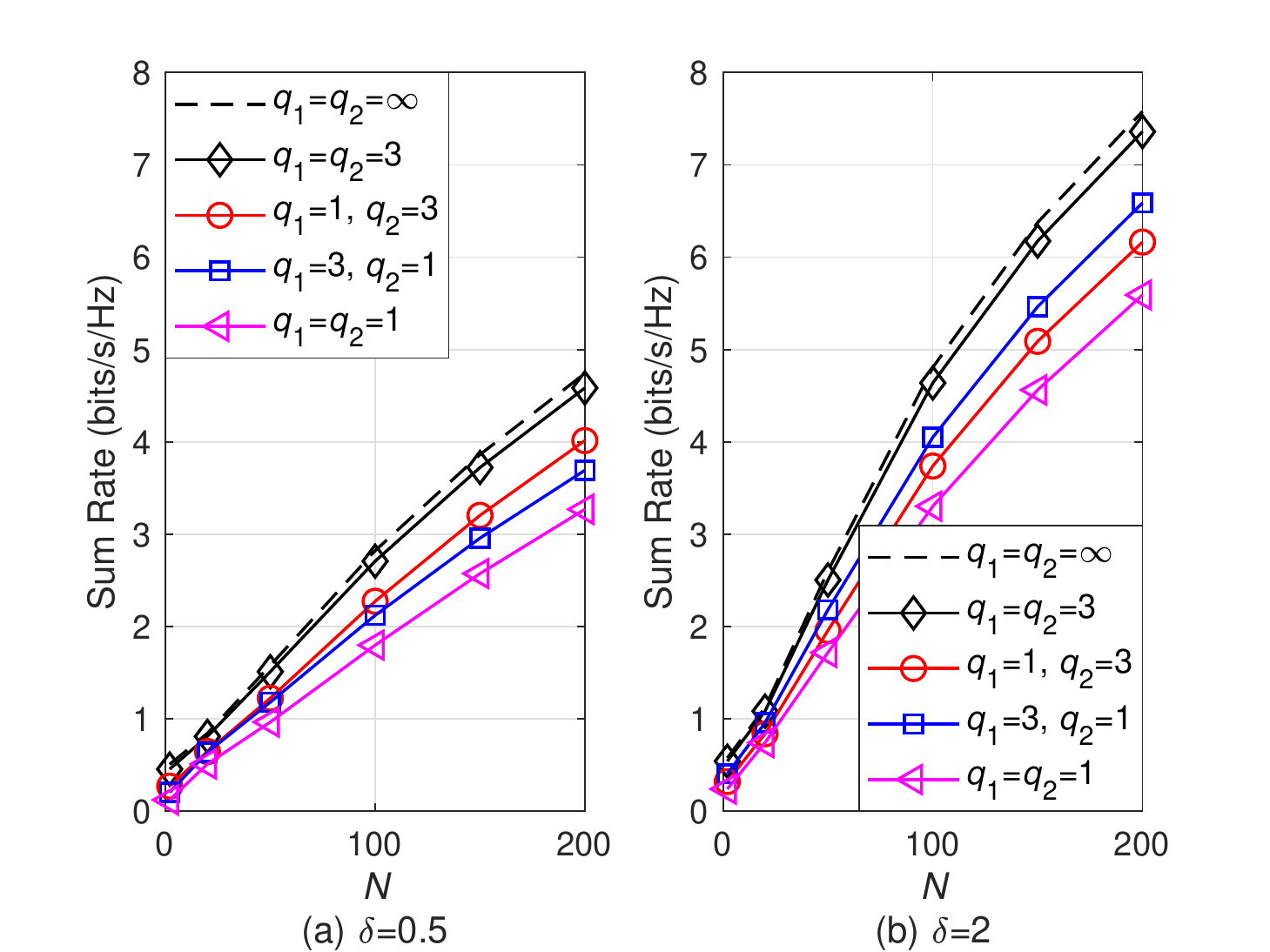}
\caption{Achievable rate performance with $r_{\textrm{R}}=r_{\textrm{B}}=0.8$ and different ADC resolution bits.}\label{impact_ADC}
\end{figure}

Finally, we turn to study the impact of ADC resolution bits at the RS and the BS in Fig.~\ref{impact_ADC}, where the spatial correlation levels are set as $r_{\textrm{R}}=r_{\textrm{B}}=0.8$ and both $\delta=0.5$ and $\delta=2$ are considered. Using three-bit ADCs at both the RS and the BS significantly improves the performance compared to using one-bit ADCs and is close to the ideal case with $q_1=q_2=\infty$. For $\delta=0.5$, respectively deploying lower and higher resolution ADCs at the RS and the BS ($q_1=1, q_2=3$) outperforms the case with $q_1=3, q_2=1$. The relationship between the two cases reverses for $\delta=2$. For the RS and the BS, the one equipped with more antennas is allowed to deploy lower resolution ADCs.



\section{Conclusion}

In this paper, we analyze the performance of massive MIMO relay systems by considering the main practical issues including spatial correlation, estimated CSI, and imperfect ADC quantization. A closed-form approximation of the achievable rate is derived. Further analysis straightforwardly reveals how the power scaling, spatial correlation, and ADC resolution impact the achievable rate performance in different system settings. The analysis guides us how to deploy the RS and the BS and adjust the numbers of antennas and ADC resolution bits at them harmoniously for the general massive MIMO relay systems.


\begin{appendix}
\subsection{Preliminary Lemma}

To pave the achievable rate analysis, we propose the following novel lemma to well depict the statistical property of the double-side correlated channels. This general lemma can be also trivially reduced to the i.i.d. or single-side correlated case.

\emph{Lemma 1:} Consider an $M$ by $N$ matrix $\mathbf{Y}=\mathbf{P}\mathbf{X}\mathbf{Q}$, where $\mathbf{P}\in \mathbb{C}^{M\times M}$ and $\mathbf{Q}\in \mathbb{C}^{N\times N}$ are deterministic matrices and $\mathbf{X}\in \mathbb{C}^{M\times N}$ is a random matrix with i.i.d. $\mathcal{CN}(0,1)$ entries. Let $\mathbf{y}_i$ be the $i$th column of $\mathbf{Y}$, then we have
\setlength{\arraycolsep}{0.0em}
\begin{eqnarray}
&&(i)\quad\mathbb{E}\left\{\mathbf{y}_i^H\mathbf{y}_j\right\}=\varrho_{ij}\textrm{tr}(\mathbf{P}^H\mathbf{P}),\label{eqn_EyiHyj}\nonumber\\
&&(ii)\;\;\,\mathbb{E}\left\{|\mathbf{y}_i^H\mathbf{y}_j|^2\right\}=\varrho^2_{ij}\textrm{tr}^2(\mathbf{P}^H\mathbf{P}) +\varrho_{ii}\varrho_{jj}\|\mathbf{P}^H\mathbf{P}\|^2_{F},\;\;\label{eqn_EyiHyj2}\nonumber\\
&&(iii)\;\,\mathbb{E}\left\{y_{mi}^{*}y_{mj}\right\}=\varrho_{ii}\zeta_{mm},\nonumber\\
&&\quad\quad\;\mathbb{E}\left\{|y_{mi}|^2|y_{mj}|^2\right\}=\zeta_{mm}^2(\varrho_{ij}^2+\varrho_{ii}\varrho_{jj}),\nonumber
\end{eqnarray}
where $\zeta_{ij}$ and $\varrho_{ij}$ denote the $(i,j)$th elements of the matrices $\mathbf{P}^H\mathbf{P}$ and $\mathbf{Q}^H\mathbf{Q}$, respectively.
\begin{IEEEproof}
We first vectorize $\mathbf{Y}$ as
\setlength{\arraycolsep}{0.0em}
\begin{eqnarray}
\label{eqn_vecY}
\bar{\mathbf{y}}=\textrm{vec}(\mathbf{Y})=(\mathbf{Q}^T\otimes\mathbf{P}) \textrm{vec}(\mathbf{X})\triangleq (\mathbf{Q}^T\otimes\mathbf{P})\bar{\mathbf{x}},
\end{eqnarray}
based on which, $\mathbf{y}_i$ is written as
\setlength{\arraycolsep}{0.0em}
\begin{eqnarray}
\label{eqn_yi}
\mathbf{y}_i=\bar{\mathbf{y}}((i-1)M+1:iM)=(\mathbf{q}_{i}^T\otimes\mathbf{P})\bar{\mathbf{x}}\triangleq \mathbf{U}_{i}\bar{\mathbf{x}},
\end{eqnarray}
where $\mathbf{q}_{i}$ is the $i$th column of $\mathbf{Q}$.

Then we have
\begin{eqnarray}
\mathbb{E}\left\{\mathbf{y}_i^H\mathbf{y}_j\right\}&&=\mathbb{E}\left\{\bar{\mathbf{x}}^H\mathbf{U}_{i}^H\mathbf{U}_{j}\bar{\mathbf{x}}\right\} \nonumber\\
&&=\textrm{tr}(\mathbf{U}_{i}^H\mathbf{U}_{j})\overset{(a)}=\textrm{tr}(\mathbf{q}_{i}^*\mathbf{q}_{j}^T\!\otimes \mathbf{P}^H\mathbf{P}),
\end{eqnarray}
\vspace{-0.3cm}
\begin{eqnarray}
\!\!\!\!\!\!\mathbb{E}\left\{|\mathbf{y}_i^H\mathbf{y}_j|^2\right\} &&=\mathbb{E}\left\{|\bar{\mathbf{x}}^H\mathbf{U}_{i}^H\mathbf{U}_{j}\bar{\mathbf{x}}|^2\right\} \nonumber\\ &&\overset{(b)}=\textrm{tr}^2(\mathbf{U}_{i}^H\mathbf{U}_{j})+\|\mathbf{U}_{i}^H\mathbf{U}_{j}\|_{F}^2 \nonumber\\
&&\overset{(a)}=\textrm{tr}^2(\mathbf{q}_{i}^*\mathbf{q}_{j}^T\!\otimes \mathbf{P}^H\mathbf{P})\!+\!\|\mathbf{q}_{i}^*\mathbf{q}_{j}^T\!\otimes \mathbf{P}^H\mathbf{P}\|_{F}^2,\,\,\,\,
\end{eqnarray}
where ($a$) is obtained based on $(\mathbf{A}\otimes \mathbf{B})(\mathbf{C}\otimes \mathbf{D})=\mathbf{A}\mathbf{C}\otimes\mathbf{B}\mathbf{D}$ and ($b$) is according to \cite[Lemma 2]{E. Bjornson}. Then we have
\setlength{\arraycolsep}{0.0em}
\begin{eqnarray}
\textrm{tr}(\mathbf{q}_{i}^*\mathbf{q}_{j}^T\otimes \mathbf{P}^H\mathbf{P}) &&=\textrm{tr}(\mathbf{q}_{i}^*\mathbf{q}_{j}^T)\textrm{tr}(\mathbf{P}^H\mathbf{P}) \nonumber\\ &&=\mathbf{q}_{i}^H\mathbf{q}_{j}\textrm{tr}(\mathbf{P}^H\mathbf{P})=\varrho_{ij}\textrm{tr}(\mathbf{P}^H\mathbf{P}),
\end{eqnarray}
\vspace{-0.3cm}
\begin{eqnarray}
\|\mathbf{q}_{i}^*\mathbf{q}_{j}^T\!\otimes \mathbf{P}^H\mathbf{P}\|_{F}^2 &&=\|\mathbf{q}_{i}^*\mathbf{q}_{j}^T\|_{F}^2 \|\mathbf{P}^H\mathbf{P}\|_{F}^2\nonumber\\
&&=\textrm{tr}(\mathbf{q}_{i}^*\mathbf{q}_{j}^T\mathbf{q}_{j}^*\mathbf{q}_{i}^T)\|\mathbf{P}^H\mathbf{P}\|_{F}^2\nonumber\\
&&=\mathbf{q}_{i}^T\mathbf{q}_{i}^*\mathbf{q}_{j}^T\mathbf{q}_{j}^*\|\mathbf{P}^H\mathbf{P}\|_{F}^2\nonumber\\
&&=\varrho_{ii}\varrho_{jj}\|\mathbf{P}^H\mathbf{P}\|^2_{F},
\end{eqnarray}
which prove ($i$) and ($ii$). ($iii$) can be proved similarly.

\end{IEEEproof}

\subsection{Proof of Theorem 1}
\allowdisplaybreaks[3]

\emph{1) Derivation of $\kappa$:} For (\ref{eqn_kappa}), we have
\setlength{\arraycolsep}{0.0em}
\begin{eqnarray}
&&\text{tr}(\mathbb{E}\{\hat{\mathbf{F}}^H\mathbf{F}\mathbf{F}^H\hat{\mathbf{F}}\}) \!=\!\text{tr}(\mathbb{E}\{\hat{\mathbf{F}}^H\hat{\mathbf{F}}\hat{\mathbf{F}}^H\hat{\mathbf{F}}\}) \!+\!\text{tr}(\mathbb{E}\{\hat{\mathbf{F}}^H\tilde{\mathbf{F}}\tilde{\mathbf{F}}^H\hat{\mathbf{F}}\})\nonumber\\
&&=\sum_{i=1}^K\biggl(\mathbb{E}\{|\hat{\mathbf{f}}_{i}^H \hat{\mathbf{f}}_{i}|^2\}\!+\!\sum_{l \neq i}^K \mathbb{E}\{|\hat{\mathbf{f}}_{i}^H \hat{\mathbf{f}}_{l}|^2\}+\sum_{j=1}^K \mathbb{E}\{|\hat{\mathbf{f}}_{i}^H \tilde{\mathbf{f}}_{j}|^2\}\biggr)\nonumber\\
&&\overset{(c)}=\!\sum_{i=1}^K\hat{\beta}_i\biggl(\hat{\beta}_i(\textrm{tr}(\hat{\mathbf{T}}_{\textrm{R}_{\textrm{r}}}))^2 \!+\!\|\hat{\mathbf{T}}_{\textrm{R}_{\textrm{r}}}\|_{F}^2\sum_{l=1}^K\hat{\beta}_l \!+\!\textrm{tr}(\tilde{\mathbf{T}}_{\textrm{R}_{\textrm{r}}}\hat{\mathbf{T}}_{\textrm{R}_{\textrm{r}}})\sum_{j=1}^K\tilde{\beta}_j\biggr),\nonumber\\
&&
\end{eqnarray}
\vspace{-0.3cm}
\begin{eqnarray}
&&\text{tr}(\mathbb{E}\{\hat{\mathbf{F}}^H\text{diag}(\mathbf{F}\mathbf{F}^H)\hat{\mathbf{F}}\}) \nonumber\\ &&=\text{tr}(\mathbb{E}\{\text{diag}(\hat{\mathbf{F}}\hat{\mathbf{F}}^H)\hat{\mathbf{F}}\hat{\mathbf{F}}^H\}) +\text{tr}(\mathbb{E}\{\text{diag}(\tilde{\mathbf{F}}\tilde{\mathbf{F}}^H)\hat{\mathbf{F}}\hat{\mathbf{F}}^H\})\nonumber\\
&&=\sum_{n=1}^{N}\sum_{i=1}^K\biggl( \mathbb{E}\{|\hat{f}_{ni}|^4\}+\sum_{l \neq i}^K \mathbb{E}\{|\hat{f}_{ni}|^2 |\hat{f}_{nl}|^2\}\nonumber\\
&&\quad+\mathbb{E}\{|\hat{f}_{ni}|^2\}\sum_{j=1}^K\mathbb{E}\{|\tilde{f}_{nj}|^2\}\biggr)\nonumber\\
&&\overset{(d)}=\!\sum_{n=1}^{N}\hat{t}_{\textrm{R}_{\textrm{r}},nn}\!\sum_{i=1}^K\hat{\beta}_i\biggl(\!\hat{t}_{\textrm{R}_{\textrm{r}},nn}\biggl(\hat{\beta}_i\!+\!\sum_{l=1}^K \hat{\beta}_l\biggr)\!+\!\tilde{t}_{\textrm{R}_{\textrm{r}},nn}\!\sum_{l=1}^K\tilde{\beta}_l\biggr)\!, \;\; 
\end{eqnarray}
\vspace{-0.3cm}
\begin{eqnarray}
\text{tr}(\mathbb{E}\{\hat{\mathbf{F}}^H\hat{\mathbf{F}}\})=\sum_{l=1}^K \mathbb{E}\{\hat{\mathbf{f}}_{l}^H \hat{\mathbf{f}}_{l}\}\overset{(e)}=\textrm{tr}(\hat{\mathbf{T}}_{\textrm{R}_{\textrm{r}}})\sum_{l=1}^K\hat{\beta}_l,
\end{eqnarray}
where ($c$), ($d$), and ($e$) are derived based on Lemma 1 ($ii$), ($iii$), and ($i$), respectively.

\emph{2) Sum Rate Approximation:}
According to \cite[Lemma 1]{Q. Zhang}, we approximate the achievable sum rate as
\begin{eqnarray}
\label{eqn_app_proof}
\!\!{R}_{\textrm{sum}}\approx\mu\sum_{k=1}^K\log_2\!\left(1+\frac{ \chi\mathbb{E}\{|\hat{\mathbf{g}}_{k}^H\hat{\mathbf{G}}\hat{\mathbf{F}}^H\hat{\mathbf{f}}_{k}|^2\}} {\mathbb{E}\left\{B_2\right\}}\!\right)\triangleq \hat{R}_{\textrm{sum}}.
\end{eqnarray}

According to \cite{P. Dong_a}, we have
\begin{eqnarray}
\label{eqn_E_gGFf2}
&&\mathbb{E}\{|\hat{\mathbf{g}}_{k}^H\hat{\mathbf{G}}\hat{\mathbf{F}}^H\hat{\mathbf{f}}_{k}|^2\}\nonumber\\
&&=\mathbb{E}\{\vert \hat{\mathbf{g}}_k^H\hat{\mathbf{g}}_k\vert^2\}\mathbb{E}\{\vert \hat{\mathbf{f}}_k^H\hat{\mathbf{f}}_k \vert^2\}+\sum_{i \neq k}^K \mathbb{E}\{\vert \hat{\mathbf{g}}_k^H\hat{\mathbf{g}}_i\vert^2\}\mathbb{E}\{\vert \hat{\mathbf{f}}_i^H\hat{\mathbf{f}}_k \vert^2\}\nonumber\\
&&\overset{(c)}=\eta^2\hat{t}_{\textrm{R}_{\textrm{t}},kk}^2\hat{\beta}_k^2\left(\textrm{tr}^2(\hat{\mathbf{T}}_{\textrm{B}_{\textrm{r}}})+\|\hat{\mathbf{T}}_{\textrm{B}_{\textrm{r}}}\|_{F}^2\right) \textrm{tr}^2(\hat{\mathbf{T}}_{\textrm{R}_{\textrm{r}}}) \nonumber\\ &&+\eta^2\hat{\beta}_k\|\hat{\mathbf{T}}_{\textrm{R}_{\textrm{r}}}\|_{F}^2\!\!\sum_{i=1}^K\hat{\beta}_i\!\left(\hat{t}_{\textrm{R}_{\textrm{t}},ki}^2\textrm{tr}^2(\hat{\mathbf{T}}_{\textrm{B}_{\textrm{r}}}) \!+\!\hat{t}_{\textrm{R}_{\textrm{t}},kk}\hat{t}_{\textrm{R}_{\textrm{t}},ii}\|\hat{\mathbf{T}}_{\textrm{B}_{\textrm{r}}}\|_{F}^2\right)\nonumber\\
\end{eqnarray}
\vspace{-0.3cm}
\begin{eqnarray}
&&\mathbb{E}\{|B_1|^2\}\nonumber\\
&&=\!\mathbb{E}\{\vert\hat{\mathbf{g}}_k^H\hat{\mathbf{G}}\hat{\mathbf{F}}^H\tilde{\mathbf{f}}_k\vert^2 \}\!+\!\mathbb{E}\{\vert\hat{\mathbf{g}}_k^H\tilde{\mathbf{G}}\hat{\mathbf{F}}^H\hat{\mathbf{f}}_k\vert^2  \}\!+\!\mathbb{E}\{\vert\hat{\mathbf{g}}_k^H\tilde{\mathbf{G}}\hat{\mathbf{F}}^H\tilde{\mathbf{f}}_k\vert^2 \}\nonumber\\
&&\overset{(c)}=\!\eta^2\tilde{\beta}_k \textrm{tr}(\tilde{\mathbf{T}}_{\textrm{R}_{\textrm{r}}}\hat{\mathbf{T}}_{\textrm{R}_{\textrm{r}}}\!)\! \sum_{i=1}^K\hat{\beta}_i\!\left(\hat{t}_{\textrm{R}_{\textrm{t}},ki}^2\textrm{tr}^2(\hat{\mathbf{T}}_{\textrm{B}_{\textrm{r}}}) \!+\!\hat{t}_{\textrm{R}_{\textrm{t}},kk}\hat{t}_{\textrm{R}_{\textrm{t}},ii}\|\hat{\mathbf{T}}_{\textrm{B}_{\textrm{r}}}\|_{F}^2\!\right)\nonumber\\
&&+\eta^2\hat{t}_{\textrm{R}_{\textrm{t}},kk}\hat{\beta}_k\textrm{tr}(\hat{\mathbf{T}}_{\textrm{B}_{\textrm{r}}}\!\tilde{\mathbf{T}}_{\textrm{B}_{\textrm{r}}})\!\! \left(\!\tilde{t}_{\textrm{R}_{\textrm{t}},kk}\hat{\beta}_k\textrm{tr}^2(\hat{\mathbf{T}}_{\textrm{R}_{\textrm{r}}})\!+\!\|\hat{\mathbf{T}}_{\textrm{R}_{\textrm{r}}}\|_{F}^2\!\! \sum_{i=1}^K\tilde{t}_{\textrm{R}_{\textrm{t}},ii}\hat{\beta}_i \!\right)\nonumber\\
&&+\eta^2\hat{t}_{\textrm{R}_{\textrm{t}},kk}\tilde{\beta}_k\textrm{tr}(\hat{\mathbf{T}}_{\textrm{B}_{\textrm{r}}}\tilde{\mathbf{T}}_{\textrm{B}_{\textrm{r}}}) \textrm{tr}(\hat{\mathbf{T}}_{\textrm{R}_{\textrm{r}}}\tilde{\mathbf{T}}_{\textrm{R}_{\textrm{r}}}) \sum_{i=1}^K\tilde{t}_{\textrm{R}_{\textrm{t}},ii} \hat{\beta}_i,
\end{eqnarray}
\vspace{-0.3cm}
\begin{eqnarray}
&&\mathbb{E}\{|\hat{\mathbf{g}}_{k}^H\mathbf{G}\hat{\mathbf{F}}^H\mathbf{f}_{j}|^2\}\nonumber\\
&&=\mathbb{E}\{|\hat{\mathbf{g}}_k^H\hat{\mathbf{G}}\hat{\mathbf{F}}^H\hat{\mathbf{f}}_j|^2 \}+\mathbb{E}\{|\hat{\mathbf{g}}_k^H\hat{\mathbf{G}}\hat{\mathbf{F}}^H\tilde{\mathbf{f}}_j|^2 \}\nonumber\\
&&+\mathbb{E}\{|\hat{\mathbf{g}}_k^H\tilde{\mathbf{G}}\hat{\mathbf{F}}^H\hat{\mathbf{f}}_j|^2 \}+\mathbb{E}\{|\hat{\mathbf{g}}_k^H\tilde{\mathbf{G}}\hat{\mathbf{F}}^H\tilde{\mathbf{f}}_j|^2 \}\nonumber\\
&&\overset{(c)}=\eta^2\hat{\beta}_j^2\left(\hat{t}_{\textrm{R}_{\textrm{t}},kj}^2\textrm{tr}^2(\hat{\mathbf{T}}_{\textrm{B}_{\textrm{r}}}) \!+\!\hat{t}_{\textrm{R}_{\textrm{t}},kk}\hat{t}_{\textrm{R}_{\textrm{t}},jj}\|\hat{\mathbf{T}}_{\textrm{B}_{\textrm{r}}}\|_{F}^2\right)\textrm{tr}^2(\hat{\mathbf{T}}_{\textrm{R}_{\textrm{r}}})\nonumber\\
&&+\eta^2\hat{\beta}_j\|\hat{\mathbf{T}}_{\textrm{R}_{\textrm{r}}}\|_{F}^2\!\sum_{i=1}^K\!\hat{\beta}_i\!\left(\hat{t}_{\textrm{R}_{\textrm{t}},ki}^2\textrm{tr}^2(\hat{\mathbf{T}}_{\textrm{B}_{\textrm{r}}}) \!+\!\hat{t}_{\textrm{R}_{\textrm{t}},kk}\hat{t}_{\textrm{R}_{\textrm{t}},ii}\|\hat{\mathbf{T}}_{\textrm{B}_{\textrm{r}}}\|_{F}^2\right)\nonumber\\
&&+\eta^2\tilde{\beta}_j \textrm{tr}(\tilde{\mathbf{T}}_{\textrm{R}_{\textrm{r}}}\hat{\mathbf{T}}_{\textrm{R}_{\textrm{r}}}) \sum_{i=1}^K\hat{\beta}_i\left(\hat{t}_{\textrm{R}_{\textrm{t}},ki}^2\textrm{tr}^2(\hat{\mathbf{T}}_{\textrm{B}_{\textrm{r}}}) +\hat{t}_{\textrm{R}_{\textrm{t}},kk}\hat{t}_{\textrm{R}_{\textrm{t}},ii}\|\hat{\mathbf{T}}_{\textrm{B}_{\textrm{r}}}\|_{F}^2\right)\nonumber\\
&&+\eta^2\hat{t}_{\textrm{R}_{\textrm{t}},kk}\hat{\beta}_j\textrm{tr}(\hat{\mathbf{T}}_{\textrm{B}_{\textrm{r}}}\tilde{\mathbf{T}}_{\textrm{B}_{\textrm{r}}}) \!\biggl(\!\tilde{t}_{\textrm{R}_{\textrm{t}},jj}\hat{\beta}_j \textrm{tr}^2(\hat{\mathbf{T}}_{\textrm{R}_{\textrm{r}}}) \!+\!\sum_{i=1}^K \tilde{t}_{\textrm{R}_{\textrm{t}},ii}\hat{\beta}_i\|\hat{\mathbf{T}}_{\textrm{R}_{\textrm{r}}}\|_{F}^2\!\biggr)\nonumber\\
&&+\eta^2\hat{t}_{\textrm{R}_{\textrm{t}},kk}\tilde{\beta}_j\textrm{tr}(\hat{\mathbf{T}}_{\textrm{B}_{\textrm{r}}}\tilde{\mathbf{T}}_{\textrm{B}_{\textrm{r}}}) \textrm{tr}(\hat{\mathbf{T}}_{\textrm{R}_{\textrm{r}}}\tilde{\mathbf{T}}_{\textrm{R}_{\textrm{r}}}) \sum_{i=1}^K\tilde{t}_{\textrm{R}_{\textrm{t}},ii} \hat{\beta}_i,
\end{eqnarray}
\vspace{-0.3cm}
\begin{eqnarray}
&&\mathbb{E}\{\|\hat{\mathbf{g}}_{k}^H\mathbf{G}\hat{\mathbf{F}}^H\|^2\} \!=\!\mathbb{E}\{\|\hat{\mathbf{g}}_{k}^H\hat{\mathbf{G}}\hat{\mathbf{F}}^H\|^2\} \!+\!\mathbb{E}\{\|\hat{\mathbf{g}}_{k}^H\tilde{\mathbf{G}}\hat{\mathbf{F}}^H\|^2\}\nonumber\\
&&\overset{(c)}=\eta^2\textrm{tr}(\hat{\mathbf{T}}_{\textrm{R}_{\textrm{r}}}) \sum_{i=1}^K\hat{\beta}_i\left(\hat{t}_{\textrm{R}_{\textrm{t}},ki}^2\textrm{tr}^2(\hat{\mathbf{T}}_{\textrm{B}_{\textrm{r}}}) +\hat{t}_{\textrm{R}_{\textrm{t}},kk}\hat{t}_{\textrm{R}_{\textrm{t}},ii}\|\hat{\mathbf{T}}_{\textrm{B}_{\textrm{r}}}\|_{F}^2\right) \nonumber\\ &&\quad+\eta^2\hat{t}_{\textrm{R}_{\textrm{t}},kk}\textrm{tr}(\hat{\mathbf{T}}_{\textrm{R}_{\textrm{r}}}) \textrm{tr}(\hat{\mathbf{T}}_{\textrm{B}_{\textrm{r}}}\tilde{\mathbf{T}}_{\textrm{B}_{\textrm{r}}})\sum_{i=1}^K\hat{\beta}_i \tilde{t}_{\textrm{R}_{\textrm{t}},ii},
\end{eqnarray}
\vspace{-0.3cm}
\begin{eqnarray}
&&\mathbb{E}\{|\hat{\mathbf{g}}_{k}^H\mathbf{G}\hat{\mathbf{F}}^H\mathbf{n}_{\textrm{q}1}|^2\} =\mathbb{E}\{\hat{\mathbf{g}}_k^H\mathbf{G}\hat{\mathbf{F}}^H \mathbf{R}_{\mathbf{n}_{\text{q}1}} \hat{\mathbf{F}}\mathbf{G}^H\hat{\mathbf{g}}_k\}\nonumber\\
&&=\alpha_1(1\!-\!\alpha_1)(P_{\textrm{U}} \mathbb{E}\{\hat{\mathbf{g}}_k^H\mathbf{G}\hat{\mathbf{F}}^H \text{diag}(\mathbf{F} \mathbf{F}^H) \hat{\mathbf{F}}\mathbf{G}^H\hat{\mathbf{g}}_k\}\nonumber\\
&&\quad+ \sigma_R^2\mathbb{E}\{\|\hat{\mathbf{g}}_{k}^H\mathbf{G}\hat{\mathbf{F}}^H\|^2\})\nonumber\\
&&\overset{(c)(d)}=\alpha_1(1\!-\!\alpha_1)\eta^2P_{\textrm{U}}\biggl[\sum_{n=1}^{N}\hat{t}_{\textrm{R}_{\textrm{r}},nn}^2\sum_{i=1}^K \hat{\beta}_i\biggl(\hat{t}_{\textrm{R}_{\textrm{t}},ki}^2\textrm{tr}^2(\hat{\mathbf{T}}_{\textrm{B}_{\textrm{r}}})\nonumber\\ &&+\hat{t}_{\textrm{R}_{\textrm{t}},kk}\left(\hat{t}_{\textrm{R}_{\textrm{t}},ii}\|\hat{\mathbf{T}}_{\textrm{B}_{\textrm{r}}}\|_{F}^2 +\tilde{t}_{\textrm{R}_{\textrm{t}},ii}\textrm{tr}(\hat{\mathbf{T}}_{\textrm{B}_{\textrm{r}}}\tilde{\mathbf{T}}_{\textrm{B}_{\textrm{r}}})\right)\biggr)\biggl( \hat{\beta}_i+\sum_{l=1}^K\hat{\beta}_l\biggr)\nonumber\\
&&+\sum_{n=1}^{N}\hat{t}_{\textrm{R}_{r},nn}\tilde{t}_{\textrm{R}_{\textrm{r}},nn}\sum_{i=1}^K\hat{\beta}_i \biggl(\hat{t}_{\textrm{R}_{\textrm{t}},ki}^2\textrm{tr}^2(\hat{\mathbf{T}}_{\textrm{B}_{\textrm{r}}}) \nonumber\\ &&+\hat{t}_{\textrm{R}_{\textrm{t}},kk}\left(\hat{t}_{\textrm{R}_{\textrm{t}},ii}\|\hat{\mathbf{T}}_{\textrm{B}_{\textrm{r}}}\|_{F}^2 +\tilde{t}_{\textrm{R}_{\textrm{t}},ii} \textrm{tr}(\hat{\mathbf{T}}_{\textrm{B}_{\textrm{r}}}\tilde{\mathbf{T}}_{\textrm{B}_{\textrm{r}}})\right)\biggr)\sum_{l=1}^K\tilde{\beta}_l\biggr]\nonumber\\
&&+\alpha_1(1\!-\!\alpha_1)\sigma_R^2\mathbb{E}\{\|\hat{\mathbf{g}}_{k}^H\mathbf{G}\hat{\mathbf{F}}^H\|^2\},
\end{eqnarray}
\vspace{-0.3cm}
\begin{eqnarray}
\mathbb{E}\left\{\left\|\hat{\mathbf{g}}_{k}\right\|^2\right\}=\mathbb{E}\left\{\hat{\mathbf{g}}_k^H\hat{\mathbf{g}}_k\right\}\overset{(e)}= \eta\hat{t}_{\textrm{R}_{\textrm{t}},kk}\textrm{tr}(\hat{\mathbf{T}}_{\textrm{B}_{\textrm{r}}}).
\end{eqnarray}
\vspace{-0.3cm}
\begin{eqnarray}
&&\mathbb{E}\{|\hat{\mathbf{g}}_{k}^H\mathbf{n}_{\textrm{q}2}|^2\} =\mathbb{E}\{\hat{\mathbf{g}}_{k}^H\mathbf{R}_{\mathbf{n}_{\text{q}2}}\hat{\mathbf{g}}_{k}\}\nonumber\\
&&=\alpha_2(1\!-\!\alpha_2)\left(\frac{P_{\textrm{R}}}{K}\mathbb{E}\{\hat{\mathbf{g}}_k^H\textrm{diag}(\mathbf{G}\mathbf{G}^{H})\hat{\mathbf{g}}_k\} \!+\!\sigma_B^2\mathbb{E}\{\|\hat{\mathbf{g}}_{k}\|^2\}\right)\nonumber\\
&&\overset{(d)}=\frac{\alpha_2(1\!-\!\alpha_2)\eta^2P_{\textrm{R}}}{K}\biggl[\sum_{m=1}^M \hat{t}_{\textrm{B}_{\textrm{\textrm{r}}},mm}^2\sum_{i=1}^K\left(\hat{t}_{\textrm{R}_{\textrm{t}},ki}^2\!+\!\hat{t}_{\textrm{R}_{\textrm{t}},kk}\hat{t}_{\textrm{R}_{\textrm{t}},ii}\right)\nonumber\\
&&+\hat{t}_{\textrm{R}_{\textrm{t}},kk}\!\sum_{m=1}^M \hat{t}_{\textrm{B}_{\textrm{r}},mm}\tilde{t}_{\textrm{B}_{\textrm{r}},mm}\!\sum_{i=1}^K\tilde{t}_{\textrm{R}_{\textrm{t}},ii}\biggr] \!+\!\alpha_2(1\!-\!\alpha_2)\sigma_B^2\mathbb{E}\{\!\|\hat{\mathbf{g}}_{k}\|^2\!\}.\nonumber\\
&&
\end{eqnarray}

Therefore, $S_{k}$, $I_{k}$, $N_{k1}$, and $N_{k2}$ can be obtained by
\begin{eqnarray}
S_{k}=\chi\mathbb{E}\{|\hat{\mathbf{g}}_{k}^H\hat{\mathbf{G}}\hat{\mathbf{F}}^H\hat{\mathbf{f}}_{k}|^2\},
\end{eqnarray}
\vspace{-0.3cm}
\begin{eqnarray}
I_{k1}=\chi\left(\mathbb{E}\{|B_1|^2\}+\sum_{j\neq k}^{K}\mathbb{E}\{|\hat{\mathbf{g}}_{k}^H\mathbf{G}\hat{\mathbf{F}}^H\mathbf{f}_{j}|^2\}\right),
\end{eqnarray}
\vspace{-0.3cm}
\begin{eqnarray}
N_{k1}\!=\!\alpha_{1}^2\alpha_{2}^2\kappa^2\sigma_{\textrm{R}}^2\mathbb{E}\{\|\hat{\mathbf{g}}_{k}^H\mathbf{G}\hat{\mathbf{F}}^H\|^2\} \!+\!\alpha_{2}^2\kappa^2\mathbb{E}\{|\hat{\mathbf{g}}_{k}^H\mathbf{G}\hat{\mathbf{F}}^H\mathbf{n}_{\textrm{q}1}|^2\},\nonumber\\
&&
\end{eqnarray}
\vspace{-0.3cm}
\begin{eqnarray}
N_{k2}=\alpha_{2}^2\sigma_{\textrm{B}}^2\mathbb{E}\{\|\hat{\mathbf{g}}_{k}\|^2\} +\mathbb{E}\{|\hat{\mathbf{g}}_{k}^H\mathbf{n}_{\textrm{q}2}|^2\}.
\end{eqnarray}

\subsection{Proof of Corollary 1}
By selecting $K$ equally spaced antennas from $N$ RS antennas for signal forwarding and considering the representative and widely used exponential correlation model \cite{S. L. Loyka} to simplify the proof, we have
\begin{eqnarray}
\label{eqn_lim_tRt_ij2}
\lim\limits_{N \to \infty}t_{\textrm{R}_{\textrm{t}},ij}^2=\lim\limits_{N \to \infty} |r_{\textrm{R}}|^{2|j-i|}=\lim\limits_{N \to \infty} |r_{\textrm{R}}|^{2\omega\frac{N}{K}}=0,
\end{eqnarray}
\begin{eqnarray}
\label{eqn_lim_tRt_ijM2}
\lim\limits_{N \to \infty}t_{\textrm{R}_{\textrm{t}},ij}^2M^2&&=\lim\limits_{N \to \infty} |r_{\textrm{R}}|^{2\omega\frac{N}{K}}M^2 =\lim\limits_{N \to \infty}\left(\delta\frac{N}{(\frac{1}{|r_{\textrm{R}}|})^{\omega\frac{N}{K}}}\right)^2\nonumber\\
&&=\lim\limits_{N \to \infty}\left(\frac{\delta K}{\omega(\frac{1}{|r_{\textrm{R}}|})^{\frac{N}{K}}\ln\frac{1}{|r_{\textrm{R}}|}}\right)^2=0,
\end{eqnarray}
for $i\neq j$, where $r_{\textrm{R}}$ denotes the correlation coefficient of adjacent antennas at the RS with $0<|r_{\textrm{R}}|<1$ and $\omega=1,\ldots,K-1$.

Then substituting $P_{\textrm{U}}=\frac{E_{\textrm{U}}}{N^a}$ and $P_{\textrm{R}}=\frac{E_{\textrm{R}}}{M^b}$ into (\ref{eqn_Sk_pCSI})$-$(\ref{eqn_Nk2_pCSI}) and dividing $\kappa^2P_{\textrm{U}}$ for all terms, we can obtain $S_{k}=\mathcal{O}(N^4)$, $I_{k}=\mathcal{O}(N^3)$, $N_{k1}=\mathcal{O}(N^{3+a})$, and $N_{k2}=\mathcal{O}(N^{3+b})$, which leads to
\begin{eqnarray}
\label{eqn_lim_gammak_proof}
\lim\limits_{N \to \infty}\gamma_{k}=
\begin{cases}
\infty, & a,b<1\\
C, & \textrm{otherwise}\\
0, & a>1 \ \text{or} \ b>1
\end{cases},
\end{eqnarray}
where $C$ is bounded and turns to the corresponding values for $a<b=1$, $b<a=1$, and $a=b=1$, respectively.

\subsection{Proof of Corollary 3}
For $S_{k}$, $I_{k}$, $N_{k1}$, and $N_{k2}$ in (\ref{eqn_Sk_pCSI})$-$(\ref{eqn_Nk2_pCSI}), we omit the terms with lower order of magnitude as $N\rightarrow\infty$ and assume $\beta_{1}=\cdots=\beta_{K}=1$ for simplicity, which leads to
\begin{eqnarray}
\lim\limits_{N \to \infty}\gamma_{k}\approx\tilde{\gamma}_{k}=\frac{\tilde{S}_{k}}{\tilde{I}_{k}+\tilde{N}_{k1}+\tilde{N}_{k2}},
\end{eqnarray}
where
\begin{eqnarray}
\label{eqn_Sk_proof}
\tilde{S}_{k}=\left(M^2\!+\!\|\mathbf{T}_{\textrm{B}_{\textrm{r}}}\|_{F}^2\right)(N^2+\|\mathbf{T}_{\textrm{R}_{\textrm{r}}}\|_{F}^2),
\end{eqnarray}
\begin{eqnarray}
\label{eqn_Ik_proof}
\tilde{I}_{k}=(K\!-\!1)[\|\mathbf{T}_{\textrm{B}_{\textrm{r}}}\|_{F}^2N^2\! +\!\|\mathbf{T}_{\textrm{R}_{\textrm{r}}}\|_{F}^2(M^2 \!+\!\|\mathbf{T}_{\textrm{B}_{\textrm{r}}}\|_{F}^2)],
\end{eqnarray}
\begin{eqnarray}
\tilde{N}_{k1}=\frac{M^2N}{\alpha_1}\biggl[(1-\alpha_1)(K+1)+\frac{\sigma_{\textrm{R}}^2}{P_{\textrm{U}}}\biggr],
\end{eqnarray}
\begin{eqnarray}
\tilde{N}_{k2}=\frac{M}{\alpha_1^2\alpha_2\kappa^2P_{\textrm{U}}}\left[(1\!-\!\alpha_2) P_{\textrm{R}}+\frac{\sigma_{\textrm{B}}^2}{\eta}\right].
\end{eqnarray}

Considering the same correlation level and exponential correlation model for the RS and the BS, the correlation coefficients
of adjacent antennas at the RS and the BS are equal, i.e., $r_{\textrm{R}}=r_{\textrm{B}}=r$, which leads to $\lim\limits_{N \to \infty}\|\mathbf{T}_{\textrm{B}_{\textrm{r}}}\|_{F}^2 =\frac{1+|r|^2}{1-|r|^2}M$ and $\lim\limits_{N \to \infty}\|\mathbf{T}_{\textrm{R}_{\textrm{r}}}\|_{F}^2 =\frac{1+|r|^2}{1-|r|^2}N$. Substitute them into (\ref{eqn_Sk_proof}) and (\ref{eqn_Ik_proof}) and denote $\tilde{\gamma}_{k}^{\mathbf{T}_{\textrm{R}_{\textrm{r}}}=\mathbf{I}_{N}}$ as
\begin{eqnarray}
\tilde{\gamma}_{k}^{\mathbf{T}_{\textrm{R}_{\textrm{r}}}=\mathbf{I}_{N}}=\frac{C_1}{C_2}.
\end{eqnarray}
Then $\tilde{\gamma}_{k}^{\mathbf{T}_{\textrm{B}_{\textrm{r}}}=\mathbf{I}_{M}}$ can be written as
\begin{eqnarray}
\tilde{\gamma}_{k}^{\mathbf{T}_{\textrm{B}_{\textrm{r}}}=\mathbf{I}_{B}}=\frac{C_1+d}{C_2+(K-1)d},
\end{eqnarray}
where $d=\frac{2|r|^2}{1-|r|^2}MN(M-N)$.

Therefore, when $\delta=1$, i.e., $M=N$, we can obtain
\begin{eqnarray}
\!\!\!\!\lim\limits_{N \to \infty}\gamma_{k}^{\mathbf{T}_{\textrm{R}_{\textrm{r}}}=\mathbf{I}_{N}} \!\approx\!\tilde{\gamma}_{k}^{\mathbf{T}_{\textrm{R}_{\textrm{r}}}=\mathbf{I}_{N}} \!=\!\tilde{\gamma}_{k}^{\mathbf{T}_{\textrm{B}_{\textrm{r}}}=\mathbf{I}_{M}} \!\approx\!\lim\limits_{N \to \infty}\gamma_{k}^{\mathbf{T}_{\textrm{B}_{\textrm{r}}}=\mathbf{I}_{M}},
\end{eqnarray}
When $\delta>1$, we have
\begin{eqnarray}
\frac{C_1+d}{C_2+(K-1)d}<\frac{C_1+d}{C_2+d}<\frac{C_1}{C_2},
\end{eqnarray}
which leads to
\begin{eqnarray}
\lim\limits_{N \to \infty}\gamma_{k}^{\mathbf{T}_{\textrm{R}_{\textrm{r}}}=\mathbf{I}_{N}} >\lim\limits_{N \to \infty}\gamma_{k}^{\mathbf{T}_{\textrm{B}_{\textrm{r}}}=\mathbf{I}_{M}}.
\end{eqnarray}
When $\delta<1$, we have
\begin{eqnarray}
\frac{C_1+d}{C_2+(K-1)d}>\frac{C_1+d}{C_2+d}>\frac{C_1}{C_2},
\end{eqnarray}
which leads to
\begin{eqnarray}
\lim\limits_{N \to \infty}\gamma_{k}^{\mathbf{T}_{\textrm{R}_{\textrm{r}}}=\mathbf{I}_{N}} <\lim\limits_{N \to \infty}\gamma_{k}^{\mathbf{T}_{\textrm{B}_{\textrm{r}}}=\mathbf{I}_{M}}.
\end{eqnarray}

\subsection{Proof of Corollary 4}

For $\kappa$ in (\ref{eqn_kappa}), we omit the terms with lower order of magnitude as $N\rightarrow\infty$ and assume $\beta_{1}=\cdots=\beta_{K}=\eta=1$ for simplicity, which leads to
\begin{eqnarray}
\lim\limits_{N \to \infty}\kappa\approx\tilde{\kappa}=\sqrt{\frac{P_{\textrm{R}}}{\alpha_1^2P_{\textrm{U}}N^2K}}.
\end{eqnarray}
The relationship between $\lim\limits_{N \to \infty}\hat{R}_{\textrm{sum}}^{\alpha_1=1}$ and $\lim\limits_{N \to \infty}\hat{R}_{\textrm{sum}}^{\alpha_2=1}$ are dependent on the values of $\tilde{N}_{k1}$ and $\tilde{N}_{k2}$, and we have
\begin{eqnarray}
&&(\tilde{N}_{k1}^{\alpha_1=1}+\tilde{N}_{k2}^{\alpha_1=1})-(\tilde{N}_{k1}^{\alpha_2=1}+\tilde{N}_{k2}^{\alpha_2=1})\nonumber\\
&&=M^2N\frac{\sigma_{\textrm{R}}^2}{P_{\textrm{U}}}+\frac{M}{\alpha_1^2\alpha_2\tilde{\kappa}^2P_{\textrm{U}}}\left[(1\!-\!\alpha_2) P_{\textrm{R}}+\sigma_{\textrm{B}}^2\right]\nonumber\\
&&\quad-\frac{M^2N}{\alpha_1}\biggl[(1-\alpha_1)(K+1)+\frac{\sigma_{\textrm{R}}^2}{P_{\textrm{U}}}\biggr] -\frac{M\sigma_{\textrm{B}}^2}{\alpha_1^2\tilde{\kappa}^2P_{\textrm{U}}}\nonumber\\
&&=MN^2K\biggl(1+\frac{\sigma_{\textrm{B}}^2}{ P_{\textrm{R}}}\biggr)\biggl(\frac{1}{\alpha_2}-1\biggr) \nonumber\\ &&\quad-M^2N\biggl(K+1+\frac{\sigma_{\textrm{R}}^2}{P_{\textrm{U}}}\biggr)\biggl(\frac{1}{\alpha_1}-1\biggr)\nonumber\\
&&\overset{(f)}\approx MN^2K\biggl(\frac{1}{\alpha_2}-1\biggr)-M^2NK\biggl(\frac{1}{\alpha_1}-1\biggr)\nonumber\\
&&\overset{(g)}=MN^2K\biggl(\frac{1}{\alpha}-1\biggr)(1-\delta),
\end{eqnarray}
where ($f$) is based on $K\gg 1$, $\sigma_{\textrm{R}}^2\ll P_{\textrm{U}}$, and $\sigma_{\textrm{B}}^2\ll P_{\textrm{R}}$ and ($g$) is obtained by setting $\alpha_1=\alpha_2=\alpha$, respectively. Then the proof is completed.

\end{appendix}



\ifCLASSOPTIONcaptionsoff
  \newpage
\fi

\end{document}